# Problems of Tax Administration and Its Impact on Budget Revenues

Marika Ormotsadze

**Abstract**

The topic under study is of crucial importance, especially for developing countries. The aim of the present paper is to study the problems in revenue administration in terms of tax revenue in Georgia and analyze foreign experience in that respect. The main question arises here - What kind of tax rates should be implemented to be able to perform both functions of the taxation – fiscal and regulatory-stimulating one. Liberal method of revenue seems an attractive one for taxpayers. According to the economic situation in Georgia, the best solution is to use the liberal method. This will help business to develop and people to find jobs. Taxation system will also benefit from that. Tax rate in Georgia amounts to 15% and is the same for everyone, regardless the size of the business. The taxation system is regarded to be proportional. As for the American and European countries, taxes there are progressive. I think the same practice should be implemented in Georgia, and not only in case of taxation.

# საგადასახადო ადმინისტრირების პრობლემები და მისი გავლენა საბიუჯეტო შემოსავლებზე

მარიკა ორმოცაძე

ანოტაცია

საკვლევად აღებული საკითხი მეტად აქტუალურია, განსაკუთრებით კი განვითარებად ქვეყნებში. ნაშრომის მიზანია გადასახადების დაბეგვრის კუთხით საგადასახადო ადმინისტრირების პრობლემების შესწავლა საქართველოში და უცხოური გამოცდილების ანალიზი. გადასახადების ლიბერალური მოდელი მომხიბვლელია გადასახადის გადამხდელთათვის. საქართველოს ეკონომიკაში შექმნილი ვითარებიდან გამომდინარე, საუკეთესო ვარიანტია, რომ იგი იბეგრებოდეს არა მკაცრი, არამედ ლიბერალური მოდელით. ამით განვითარდება ბიზნესი, დასაქმდება მოსახლეობა, გაფართოვდება საგადასახადო ბაზაც. საქართველოში მოგების გადასახადის განაკვეთი 15%-ია და იგი საერთოა ყველასათვის, ყველა დარგისთვის და ყველა სიდიდის საწარმოსთვის, ანუ ითვლება რომ პროპორციულია. რაც შეეხება ამერიკის და ევროპის ქვეყნებს, თითქმის ყველა ქვეყანაში მოგების გადასახადი პროგრესულია. იგივე უნდა ხდებოდეს საქართველოშიც, და არა მხოლოდ მოგების გადასახადის შემთხვევაში.

## შესავალი

საბიუჯეტო-საგადასახადო პოლიტიკა ესაა სახელმწიფო ხელისუფლების ღონისძიებათა ერთობლიობა სახელმწიფოს მიერ გასაწევი ხარჯებისა და გადასახადების ცვლილებისა, რომელმაც უნდა უზრუნველყოს ქვეყნის მთლიანი ეროვნული პროდუქტისა და ეროვნული შემოსავლების ზრდა მოსახლეობის მაქსიმალური დასაქმების პირობებში. საგადასახადო-საბიუჯეტო სისტემა ეკონომიკის მართვის ერთერთი მნიშვნელოვანი ბერკეტია, რომლითაც სახელმწიფო ქვეყნის ეკონომიკას არეგულირებს.

სახელმწიფო ეკონომიკის ძლიერება დამოკიდებულია იმ ეკონომიკურ პოლიტიკაზე, რომელსაც მთავრობა აწარმოებს გარკვეულ პერიოდში. ქვეყნის ეკონომიკური სისტემის გამართული და ეფექტური ფუნქციონირება მნიშვნელოვნადაა დამოკიდებული სახელმწიფოს მიერ გატარებულ საგადასახადო პოლიტიკასა და საგადასახადო სისტემის გამართულ მუშაობაზე. საგადასახადო სისტემის სრულყოფასთან დაკავშირებული საკითხები ყოველთვის იყო და დარჩება ქვეყნის ეკონომიკური პოლიტიკის უმნიშვნელოვანეს საკვანძო საკითხად.

სახელმწიფოს მიერ გატარებული სწორი საგადასახადო - საბიუჯეტო პოლიტიკა ქვეყნის არსებობისა და განვითარების მყარ საფუძველს წარმოადგენს. საბაზრო ეკონომიკის პირობებში საბიუჯეტო-საგადასახადო პოლიტიკის ძირითად მიზანს ბიუჯეტის როლის გადიდება, საბიუჯეტო შემოსავლების ზრდა და მისი განაწილება-გადანაწილების ეფექტიანი მექანიზმის შემუშავება წარმოადგენს, რაც, საბოლოო ანგარიშით, უზრუნველყოფს საბიჯეტო-საგადასახადო სისტემის თანამედროვე საბაზრო ეკონომიკის მოთხოვნის შესაბამისად გარდაქმნას.

მთავრობის მიერ გატარებულ ეკონომიკურ ღონისძიებებს, რომლებიც განსაზღვრული მიზნის მისაღწევად არეგულირებენ ერთობლივ მოთხოვნას და ერთობლივ მიწოდებას, ეკონომიკური პოლიტიკა ეწოდება.

ქვეყნის ეკონომიკური პოლიტიკის შემადგენელი ნაწილია ფინანსური პოლიტიკა. იგი არის სახელმწიფოს მიერ თავისი ფუნქციებისა და ამოცანების განხორციელების უზრუნველსაყოფად შემუშავებული ღონისძიებების ერთობლიობა, რაც ვლინდება ფინანსური რესურსების მობილიზაციასა და ქვეყნის პოლიტიკური, ეკონომიკური და სოციალური სიტუაციებიდან გამომდინარე, საზოგადოების წევრებს შორის ეროვნული ეკონომიკის, პრიორიტეტულ სტრუქტურულ დარგებსა და ქვეყნის ტერიტორიულ ერთეულებს შორის განაწილება-გამოყენებაში. სახელმწიფოს ფინანსური პოლიტიკა მოიცავს მისი საქმიანობის ორ ურთიერთდაკავშირებულ სფეროს, როგორიცაა: საგადასახადო დაბეგვრის სფერო და სახელმწიფო ხარჯების სტრუქტურის რეგულირება, რითაც ხდება ეკონომიკაზე ზემოქმედება (ფისკალური პოლიტიკა) და ბიუჯეტის რეგულირება (საბიუჯეტო პოლიტიკა).

ეკონომიკის სტაბილიზაცია არის ხელისუფლების უმნიშვნელოვანესი ამოცანა. მთავრობის ფუნქციად ყოველთვის ითვლებოდა ფულად-საკრედიტო და საბიუჯეტო-საგადასახადო პოლიტიკის შემუშავება და გატარება. უნდა ითქვას, რომ ინდუსტრიულად განვითარებული ქვეყნების მაკროეკონომიკური პოლიტიკის შემუშავებასა და რეალიზაციას საფუძვლად დაედო ეკონომიკის საყოველთაო რეგულირების გამოცდილება ყოფილ საბჭოთა კავშირში. დასავლეთის ეკონომისტებმა ბევრი რამ გადაიღეს საბჭოური დაგეგმვიდან, კერძოდ ლეონტიევისეული დარგთაშორისი ბალანსის გამოყენება. ეს იყო გადასვლა აქტიურ ეკონომიკურ პოლიტიკაზე.

თანამედროვე პრაქტიკა ცხადყოფს, რომ ეფექტურია ის საგადასახადო პოლიტიკა, რომელიც ასტიმულირებს დაზოგვას და ხელს უწყობს ეკონომიკურ ზრდას. გამომდინარე აქედან, საგადასახადო პოლიტიკის უმთავრესი ამოცანა ხდება საგადასახადო ტვირთის ოპტიმალური გადანაწილება პროდუქციის მწარმოებლებსა და მომხმარებლებს შორის. ამასთან, ეკონომიკური ზრდის მიღწევის ინტერესებიდან გამომდინარე, მაქსიმალურად უნდა შეუმსუბუქდეს ზეწოლა მწარმოებლებს. ამ ამოცანით საგადასახადო პოლიტიკა გამოდის ეკონომიკის რეგულირების ინსტრუმენტის როლში.

სახელმწიფო საგადასახადო სისტემის სტრატეგიული მიზანი ბიუჯეტის დეფიციტის ეფექტიანი მართვა და ეკონომიკური აქტივობის სტიმულირებაა.

სახელმწიფო საგადასახადო პოლიტიკის მოქნილობა რეალურ საფინანსო-ეკონომიკურ გარემოში, რამდენიმე მიმართულებით შეიძლება გამოვლინდეს, კერძოდ:
) საშუალოვადიან პერიოდში ბიუჯეტის შემოსავლების პროგნოზირების პროცესში;
) საგადასახადო რეგულირების განხორციელებისას და საგადასახადო სიმძიმის გადანაწილებისას;

რაციონალური საგადასახადო სისტემის აგება უშუალოდა დაკავშირებული დასაბეგრ ბაზასთან და გადასახადის განაკვეთთან, ხოლო საგადასახადო განაკვეთების გრადაცია წარმოადგენს საგადასახადო რეგულირების მნიშვნელოვან ნაწილს, რაც აუცილებელია ბიზნესი და ბიუჯეტის სფეროში ინტერესების თანხვედრისთვის, ამიტომ გასათვალისწინებელია მსოფლიოში არსებული პრაქტიკა ცალკეული გადასახადების დიფერენცირებულობასა და პროგრესულობასთან დაკავშირებით. მაგალითად: დამატებითი ღირებულების გადასახადი დიდ ბრიტანეთში, გერმანიაში, კანადასა და მრავალ სხვა ქვეყანაში დიფერენცირებულია, როგორც პირველადი მოთხოვნილების საქონელსა და ფასწარმოქმნაზე, ისე მოსახლეობის შემოსავლებზე და ანელებს ინფლაციურ პროცესებს. საშემოსავლო გადასახადი პროგრესულია თითქმის ყველა ქვეყანაში, რაც ერთის მხრივ დადებითად აისახება ბიუჯეტის შემოსავლების ზრდაზე, ხოლო მეორე მხრივ ხელს უწყობს მოთხოვნის სტიმულირებას. მოგების გადასახადი პროგრესულია აშშ-ში, რაც ხელს უწყობს მცირე და საშუალო საწარმოების დაბალი საგადასახადო განაკვეთით დაბეგვრას.

სახელმწიფო საგადასახადო მენეჯმენტის დაბალ ეფექტიანობაზე მეტყველებს საქართველოში არსებული ბიუჯეტთაშორისი ტრანსფერების პრაქტიკაც, რადგან ამ შემთხვევაში ადგილობრივი თვითმართველობის ორგანოები არ არიან დაინტერესებული იზრუნონ თავიანთი საგადასახადო პოტენციალის ზრდაზე, ვინაიდან როგორც არასაკმარისი რესურსების მქონე ორგანიზაციები, მაინც ღებულობენ ტრანსფერებს [45-47, 52].

ოპტიმალური საგადასახადო განაკვეთის დაწესება სახელმწიფო ხარჯების ოპტიმიზაციის მიზნით, დაკავშირებულია ეკონომიკურ დასაბუთებასთან. საგადასახადო რეგულირების ღონისძიებათა რეალიზაციის მეთოდების მიხედვით, სახელმწიფო ზემოქმედება ეკონომიკაზე იყოფა ორ ურთიერთ დაკავშირებულ სფეროებად: საგადასახადო შეღავათების ერთობლიობა და საგადასახადო სანქციები. ამ სფეროების ოპტიმალური კომბინაცია უზრუნველყოფს საგადასახადო დაბეგვრის მოქნილობას და საბოლოო ჯამში საგადასახადო პოლიტიკის შედეგიანობას.

სახელმწიფო საგადასახადო მენეჯმენტის ეფექტიანობის თვალსაზრისით, მნიშვნელოვანია ასევე ინფლაციური გადასახადის განგარიშება, რომელიც შესაძლებელს ხდის დადგინდეს საგადასახადო ტვირთის სიმძიმე ცვალებადი საგადასახადო კანონმდებლობის პირობებში.

საერთაშორისო პრაქტიკის დანერგვა სახელმწიფო საგადასახადო მენეჯმენტის სფეროში მნიშვნელოვნად გააუმჯობესებს ქვეყანაში არსებულ საგადასახადო გარემოს და ხელს შეუწყობს ეკონომიკურ ზრდას.

სამწუხაროდ, დღევანდელ საქართველოში საგადასახადო პოლიტიკის და სისტემის ფორმირება მიმდინარეობს მისი სათანადო თეორიული საფუძვლების დაუმუშავებლობის პირობებში, რის გამოც, მოქმედი საგადასახადო სისტემა ხშირ ცვლილებას განიცდის, რაც ხელს უშლის ეკონომიკის სწრაფ ზრდას. ყველაზე დიდი სირთულეს ქმნის ის გარემოება, რომ არაა შემუშავებული მწარმოებლებს და მომხმარებლებს შორის საგადასახადო ტვირთის გადანაწილების ოპტიმალური მოდელი. საქართველოს თანამედროვე საგადასახდო პოლიტიკას თეორიულ საფუძვლად უნდა დაედოს კონცეფცია, რომლის არსი მდგომარეობს შემდეგში: საგადასახადო განაკვეთის გარკვეულ ზღვრამდე (30-40%) ზრდა იწვევს საგადასახადო შემოსავლების ზრდას, ხოლო მის ზევით გადიდება ამუხრუჭებს შემოსავლების გადიდებას და ეკონომიკურ პროგრესს. ამ კონცეფციის მთავარი იდეა მდგომარეობს იმაში, რომ მაღალი საგადასახადო წნეხი ახშობს სამეწარმეო და საინვესტიციო აქტივობას, რის გამოც, ეცემა წარმოების ზრდის ტემპები, ვიწროვდება საგადასახადო ბაზა და კლებულობს საგადასახადო შემოსავლების მოცულობა [43, 44, 48-50].

1. **საქართველოს საბიუჯეტო-საგადასახადო სისტემის ფორმირების თავისებურებები თანამედროვე ეტაპზე**

*1.1 საბიუჯეტო-საგადასახადო სისტემის ფორმირების თეორიული საფუძვლები*

გადასახადების წარმოშობას დიდი ხნის ისტორია აქვს. ისინი შემოღებულ იქნა ჯერ კიდევ საზოგადოებრივი ცივილიზაციის გარიჟრაჟზე, როგორც კი გაჩნდა მათზე პირველი საზოგადოებრივი მოთხოვნილება. რაც შეეხება საგადასახადო სისტემას, იგი შედარებით მოგვიანებით ჩამოყალიბდა, მაშინ როდესაც, წარმოიშვა სახელმწიფო და ფორმირდებოდა სახელმწიფოს განვითარების პარალელურად.

გადასახადების აკრების ფორმებისა და მეთოდების განვითარებამ მსოფლიოში, სამი ძირითადი ეტაპი განვლო. განვითარების საწყის ეტაპზე ძველი მსოფლიოდან შუა საუკუნეების დასაწყისამდე სახელმწიფოს არ გააჩნდა ფინანსური აპარატი გადასახადების აკრებისათვის, მას შეეძლო განესაზღვრა მხოლოდ შემოსავლების საერთო სიდიდე რომლის მირებაც სურდა, ხოლო აკრება ევალებოდა ქალაქს. მეორე ეტაპზე (XVI-XIX სს), წარმოიშვა სახელმწიფო დაწესებულებები, მათ შორის ფინანსური და სახელმწიფომ დაბეგვრასთან დაკავშირებული ფუნქციების ნაწილი საკუთარ თავზე აიღო: ადგენდა დაბეგვრის სიდიდეს და ყურადღებას აქცევდა გადასახადების აკრების პროცესს. მესამე ანუ თანამედროვე ეტაპზე სახელმწიფო უკვე თავისთავზე იღებს ყველა იმ ფუნქციას, რაც დაკავშირებულია გადასახადის დაწესებასთან, შეცვლასთან, დაბეგვრის წესების დადგენასთან, ადმინისტრირებასთან და ა.შ.

ახალი ისტორიის საწყის ეტაპზე, ევროპაში თანამედროვე სახელმწიფოები გამოჩნდა XVI-XVII საუკუნეებში სადაც ძირითადი გადასახადებიდან ცნობილი იყო მიწის გადასახადი, სულადობრივი გადასახადი, აქციზი, საბაჟო გადასახადი, კომუნალური და ადგილობრივი გადასახადები.

საქართველოს საგადასახადო სისტემას საფუძველი ჩაეყარა მისი სახელმწიფოებრივი ჩამოყალიბების პროცესში და იმ ეკონომიკურ წყობილებას ემსახურებოდა, რომელშიც მოცემულ ეტაპზე ცხოვრობდა ქვეყანა. აქედან გამომდინარე, სახელმწიფოს განვითარების სხვადასხვა ეტაპზე საგადასახადო სისტემის წინაშე კონკრეტული ამოცანები იდგა.

საქართველოს თანამედროვე საგადასახადო სისტემა ძირითადში 1991-1992 წლების მიჯნაზე, ქვეყანაში მიმდინარე კარდინალური ეკონომიკური გარდაქმნებისა და საბაზრო ურთიერთობებზე გადასვლის პერიოდში ჩამოყალიბდა [38-42]. რეალური საგადასახადო ურთიერთობების სამართლებრივი რეგულირების სამართლებრივი უქონლობა, კანონმდებლობის შესამუშავებლად დაშვებული შემჭიდროებული ვადები, ეკონომიკური და სოციალური კრიზისი ქვეყანაში, - ყველაფერმა ამან უშუალოდ იმოქმედა მის ჩამოყალიბებაზე. საქართველოში არსებული საგადასახადო სისტემა საზღვარგარეთის ქვეყნების გამოცდილების ბაზაზე იქმნებოდა. სწორედ ამის გამო იგი საერთო სტრუქტურითა და აგებულების პრინციპებით ძირითადში შეესაბამება მსოფლიო ეკონომიკაში გავრცელებულ გადასახადებით დაბეგვრის სისტემებს. ამასთან, აღსანიშნავია ის გარემოებაც, რომ საქართველოს საგადასახადო სისტემის ძირითადი ელემენტები ყალიბდებოდა ეროვნული სპეციფიკის გათვალისწინებით.

საქართველოში დამოუკიდებელი საგადასახადო სისტემის ჩამოყალიბება გასული საუკუნის 90-იან წლებიდან დაიწყო. იმ პერიოდში საგადასახადო სისტემის ძირითადი მარეგულირებელი ნორმატიული აქტი იყო საქართველოს რესპუბლიკის კანონი „საგადასახადო სისტემის საფუძვლების შესახებ" 1997 წლის 13 ივნისს საქართველოს პარლამენტის მიერ მიღებული იქნა საგადასახადო კოდექსი. საქართველოს საგადასახადო კოდექსმა განსაზღვრა საგადასახადო სისტემის სტრუქტურა და ფუნქციონირების ზოგადი მექანიზმი, გადასახადის სახეები, მათი გამოანგარიშებისა და გადახდის წესი, საგადასახადო ორგანოებისა და გადასახადის გადამხდელების ვალდებულებები და პასუხისმგებლობა, საგადასახადო-სამართლებრივი პრინციპები, საგადასახადო ტერმინოლოგია და ა.შ. კოდექსმა და საგადასახადო სისტემამ მრავალი ცვლილება განიცადა და ამჟამად ჩვენს ქვეყანაში მოქმედი გადასახადები მათი რაოდენობის, გადასახადების გამოანგარიშების წესის სიმარტივისა და გადასახადის გადამხდელის უფლებების დაცვის მხრივ ერთ-ერთი მიმზიდველი და შთამბეჭდავია პოსტსოციალისტურ სახელმწიფოებში.

საგადასახადო კოდექსში გამოიკვეთა შემდეგი ძირითადი მომენტები:
) შემცირდა საერთო სახელმწიფოებრივი გადასახადების რიცხვი;
) გამოიკვეთა გადასახადის დაწესებისა და გადახდევინების სამართლებრივი საფუძვლები;
) აიკრძალა კოდექსით გაუთვალისწინებელი გადასახადების ვინმესთვის დაკისრება;
) განიმარტა გადასახადის ცნება, რომელშიც გამოიკვეთა მისი გადახდის აუცილებლობა;
) ჩამოყალიბდა საქმიანობის სახეები, სამეწარმეო და არასამეწარმეო, ეკონომიკური საქმიანობა, საქველმოქმედო და რელიგიური საქმიანობა, დაქირავებით მუშაობა;
) საგადასახადო კოდექსი დადგა გადასახადების საკითხში ყველა ნორმატიულ აქტს მაღლა;
) განიმარტა საწარმოს, ინდივიდუალური საწარმოს, ორგანიზაციის, საბიუჯეტო, საქველმოქმედო და რელიგიური ორგანიზაციების ტერმინები;
) შემოდებულ იქნა ერთობლივი შემოსავლებიდან ხარჯების გამოქვითვის მექანიზმი და საგადასახადო აღრიცხვის წესები;

თეორიული თვალსაზრისით, ურთიერთობები დაბეგვრის სფეროში წარმოადგენს საფინანსო ურთიერთობების შემადგენელ ნაწილს და საგადასახადო სისტემა ორგანულად უკავშირდება საფინანსო-საბიუჯეტო სისტემას. სახელმწიფოს საგადასახადო პოლიტიკა კი წარმოადგენს ქვეყნის საფინანსო პოლიტიკის განუყოფელ ნაწილს.

გადასახადები დაწესებულია სახელმწიფო ხარჯების დაფარვის უზრუნველყოფის მიზნით და იგი წარმოადგენს სახელმწიფო ბიუჯეტის შევსების ერთ-ერთ ძირითად წყაროს.

გადასახადის გადახდა ბიუჯეტში ხორციელდება ფულადი ფორმით. საბიუჯეტო კანონმდებლობის თანახმად, სახელმწიფო ხაზინაში გადასახადების გარდა ირიცხება სხვა სახის შენატანებიც. მაგალითად, ნებართვების, ლიცენზიების გაცემისა და სხვა სახის სახელმწიფო მომსახურების გაწევიდან მიღებული საფასური, სახელმწიფო ქონების იჯარით გაცემიდან მიღებული შემოსავლები, როიალტი, შემოსავლები სახელმწიფო ქონების რეალიზაციიდან და ა.შ.

სახელმწიფო ბიუჯეტი შედგება ორი ურთიერთდაკავშირებული ნაწილისგან : 1. საშემოსავლო ნაწილი, რომელიც მოიცავს ფულადი სახსრების შემოსავლების ჩამონათვალს და 2. გასავლის ნაწილი, რომელიც აერთიანებს სახელმწიფოს მიერ გასაწევ ყველა ხარჯს.

ბიუჯეტი არის ქვეყნის პოლიტიკური და სოციალურ-ეკონომიკური არსის ყველაზე უკეთ გამოსახვის საშუალება. ამ შესაძლებლობებს, მას ანიჭებს მასში შემავალი შემოსულობები, მასზე დაკისრებული სოციალური და ეკონომიკური მოვალეობები.

ბიუჯეტის არსსა და შინაარს ყველაზე ნათლად გამოხატავს მის მიერ განხორციელებული ფუნქციები.

ბიუჯეტი ასრულებს ოთხ ფუნქციას, ესენია:

1. ერთობლივი ეროვნული პროდუქტისა და ეროვნული შემოსავლის გადანაწილება;
2. ეკონომიკის სახელმწიფოებრივი რეგულირება;
3. სოციალური პოლიტიკის ფინანსური უზრუნველყოფა;
4. ფულადი სახსრების ცენტრალიზებული ფონდების ფორმირებასა და გამოყენებაზე კონტროლი; [18, გვ106]

ეკონომიკის განვითარებაზე ზემოქმედებს ბიუჯეტის მეშვეობით განაწილებული ერთობლივი ეროვნული პროდუქტისა და ეროვნული შემოსავლის მოცულობის ზრდა. ამ დროს ძლიერდება ბიუჯეტის გავლენა სახელმწიფოს პოლიტიკური და სოციალურ-ეკონომიკური საქმიანობის წარმართვაზე.  იმისათვის, რომ თავიდან ავიცილოთ წარსულში ეკონომიკაში არსებული მთელი რიგი ნაკლოვანებები და არასასურველი შედეგები, მიმაჩნია რომ,    ბიუჯეტის პრიორიტეტს უნდა წარმოადგენდეს მოსახლეობის სხვადასხვა მოთხოვნილებებისა და სახელმწიფოს სრულყოფილი ფუნქციონირებისათვის აუცილებელი ფულადი სახსრების მობილიზება, გამოყენება და ხარჯვაზე კონტროლი.

ბიუჯეტი არის სახელმწიფოს ფინანსური რესურსების ძირითადი ბაზა, რომლის ფორმირებაც გადასახადების მეშვეობით წარმოებს. ამიტომ, სახელმწიფოს მხრიდან გატარებული უნდა იყოს სწორი საგადასახადო პოლიტიკა, უნდა იზრუნოს მისი სწორი მიმართულებით წარმართვაზე, რაც ქვეყანაში ეკონომიკური კრიზისის წარმოშობის რისკის შემცირების წინაპირობაა.

დამოუკიდებელი საქართველო, რომელიც ეკონომიკური რეფორმების საწყის ეტაპზე იმყოფებოდა, ვერ აკმაყოფილებდა საბაზრო ეკონომიკის მოთხოვნებს, პრაქტიკულად არ გააჩნდა ბიუჯეტი და ძირითადად საზღვარგარეთიდან მიღებული დახმარებებით საზრდოობდა. წლების განმავლობაში უგულებელყოფილი იყო ბიუჯეტის ფორმირების თეორიულ-მეთოდოლოგიური საფუძვლები და ბიუჯეტის ფორმირება ხდებოდა სტრატეგიული მიზნებისა და ამოცანების გათვალისწინების გარეშე, რის გამოც სახელმწიფო ვერ ახდენდა ეკონომიკის სტაბილიზაციაზე დადებით ზეგავლენას.

საბაზრო ეკონომიკის მოთხოვნების დაკმაყოფილებისთვის მნიშვნელოვანი ნაბიჯი გადაიდგა 1996 წლის 29 მაისს, როდესაც საქართველოს პარლამენტმა მიიღო კანონი „საბიუჯეტო სისტემისა და საბიუჯეტო უფლებამოსილებათა შესახებ", სადაც ბიუჯეტი წარმოდგენილი იყო, როგორც სახელმწიფო ხელისუფლების ორგანოებისათვის საჭირო ფულადი სახსრების მობილიზაციისა და გამოყენების ძირითადი ფინანსური გეგმა. [10]

2003 წელს დამტკიცდა „საქართველოს საბიუჯეტო სისტემის შესახებ" კანონი. ახლად მიღებული კანონით ძველისგან განსხვავებით ერთმანეთისგან გაიმიჯნა ცენტრალური ბიუჯეტი, სახელმწიფო ბიუჯეტი, ავტონომიური რესპუბლიკების ბიუჯეტები, ადგილობრივი ბიუჯეტი და სპეციალური ფონდები.[31]

კანონის თანახმად ცენტრალური ბიუჯეტი იყო ცენტრალური ხელისუფლების ბიუჯეტი, რომელიც მოიცავს საკანონმდებლო, აღმასრულებელი და სასამართლო ხელისუფლების შემოსულობებს, ხოლო სახელმწიფო ბიუჯეტი – ცენტრალური ბიუჯეტისა და სპეციალური ფონდების ერთობლიობა. [36, თავი I, მუხლი 3] ზემოაღნიშნული კანონით მოხდა შემოსავლების და ხარჯების პრინციპების გამიჯვნა. დროთა განმავლობაში კანონმა განიცადა მრავალი სასიკეთო ცვლილება. ავტონომიური რესპუბლიკებისა და საქართველოს სხვა ტერიტორიულ ერთეულებს უფლება მიეცათ, რომ ბიუჯეტში მთლიანად ჩარიცხულიყო: მიწის გადასახადი, ქონების და სხვა საერთო სახელმწიფოებრივი გადასახადები, აგრეთვე სამეწარმეო საქმიანობის, სასტუმროს, რეკლმის და სხვა გადასახადები.

2010 წლის 1 იანვრიდან ამოქმედდა საქართველოს საბიუჯეტო კოდექსი, რაც უმნიშვნელოვანეს ნაბიჯად ითვლება სახელმწიფო ბიუჯეტის ფორმირების მიმართულებით. ქვეყნის საბიუჯეტო სისტემა, მოქმედი კანონმდებლობის შესაბამისად ჩამოყალიბდა: ქვეყნის სახელმწიფო ბიუჯეტის, ავტონომიური რესპუბლიკის ბიუჯეტებისა და ადგილობრივი თვითმართველი ერთეულების ბიუჯეტების ეთობლიობის სახით.

მოქმედი საბიუჯეტო კოდექსის საფუძველზე შემოდებული იქნა საქართველოს ნაერთი ბიუჯეტისა და ავტონომიური რესპუბლიკების ნაერთი ბიუჯეტების ცნება. საქართველოს ნაერთი ბიუჯეტი მოიცავს ქვეყნის სახელმწიფო, ავტონომიური რესპუბლიკების რესპუბლიკური და თვითმართველი ერთეულების ბიუჯეტებს, ხოლო ავტონომიური რესპუბლიკების ნაერთი ბიუჯეტი მოიცავს აფხაზეთისა და აჭარის ავტონომიური რესპუბლიკების რესპუბლიკურ და მის შემადგენლობაში შემავალი ადგილობრივი თვითმართველი ერთეულების ბიუჯეტებს.

## 1.2 საქართველოს საბიუჯეტო-საგადასახადო პოლიტიკის ფორმირების თავისებურებები

საბიუჯეტო-საგადასახადო პოლიტიკის ძირითად მიზანს ბიუჯეტის როლის ზრდა, ბიუჯეტის შემოსულობების და გადასახდელების გონივრული განაწილება და საბიუჯეტო-საგადასახადო სისტემის საბაზრო ეკონომიკის მოთხოვნის შესაბამისად ფორმირება-ფუნქციონირება წარმოადგენს. ეკონომიკაზე ზემოქმედება და ბიუჯეტის რეგულირება ხდება საგადასახადო დაბეგვრის სფეროს და სახელმწიფო ხარჯების სტრუქტურის რეგულირებით.

ხელისუფლების მიერ შემუშავებულ საგადასახადო სტრატეგიაზე დამოკიდებულია ეკონომიკის ზრდა, ეროვნული წარმოების სტიმულირება და საინვესტიციო გარემოს ფუნქციონირება. ქვეყნის, საბიუჯეტო-საგადასახადო პოლიტიკის მნიშვნელოვან ამოცანას წარმოადგენს ბიუჯეტის საშემოსავლო ნაწილის ზრდა და მისი განაწილების ეფექტური მექანიზმის შემუშავება.

სახელმწიფო ბიუჯეტში შემოსავლების მიღება და სხვადასხვა ღონისძიებების დაფინანსება ხორციელდება კონკრეტული სქემის მიხედვით. საქართველოს ფინანსთა სამინისტროს სახაზინო სამსახურის მეშვეობით, რომელსაც ეროვნულ ბანკში გახსნილი აქვს შემოსავლების ერთიანი ანგარიში. სახელმწიფო ხაზინა პასუხისმგებელია სახელმწიფო, აფხაზეთისა და აჭარის ავტონომიური რესპუბლიკების და ტერიტორიული ერთეულების ბიუჯეტების შემოსულობათა და გადასახადების სრულ და სწორ აღრიცხვაზე.

საბიუჯეტო თანხების მოძრაობის სქემამ რამდენჯერმე განიცადა ცვლილებები. დღეისათვის მოქმედებს სქემა, რომლის მიხედვითაც გადამხდელი გადასახადს იხდის რაიონის კომერციული ბანკის ფილიალში, კომერციული ბანკის ფილიალიდან შემოსავლები გადაეცემა მის სათაო ოფისს, ხოლო შემდეგ აღნიშნული თანხები მიემართება ეროვნულ ბანკში. ეროვნულ ბანკში განთავსებულია ხაზინის ერთიანი ანგარიშში სადაც თავს იყრის მთელი თანხები. ერთიანი ანგარიშში ქვეანგარიშებია საგადასახადო და საბაჟო დეპარტამენტის სარეზერვო ფონდები და დეპოზიტები. (იხ.სქემა №1) რაც შეეხება ხარჯების მოძრაობას, სახაზინო სამსახურის მიერ ხარჯების გაწევა სამი ეტაპისგან შედგება: ვალდებულების აღება, დამოწმება და სახსრების გადახდა. საბიუჯეტო ორგანიზაცია ბიუჯეტით დამტკიცებული სახსრების მისარებად სახელმწიფო ხაზინაში წარადგენს შემდეგ დოკუმენტაციას: ვალდებულების დოკუმენტი, დამოწმების დოკუმენტი, საგადასახადო დავალება. (იხ.სქემა №2) [25] საქართველოს საბიუჯეტო სისტემის სრულყოფა მნიშვნელოვნად არის დამოკიდებული საბიუჯეტო სისტემაში შემავალი სხვადასხვა დონის ბიუჯეტებს შორის სახსრების მობილიზაციისა და გადანაწილების ეფექტური მექანიზმის შემუშავებაზე.

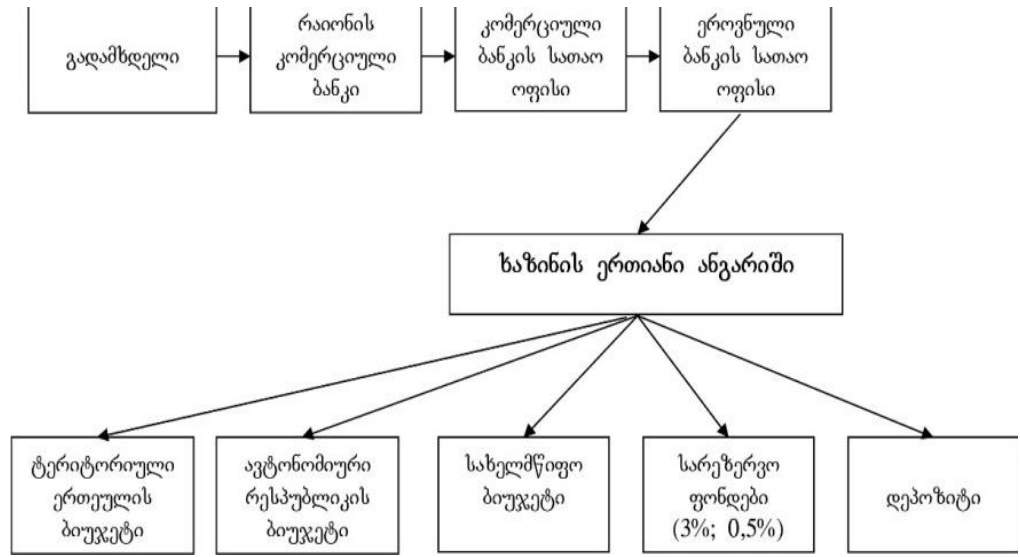

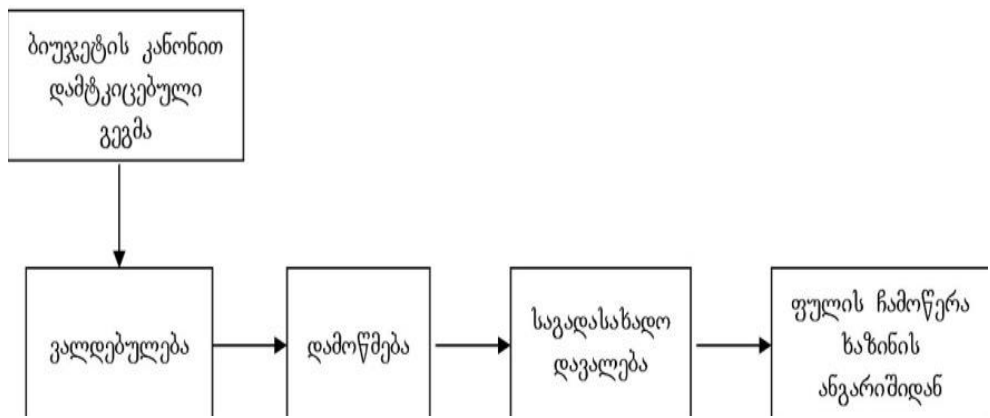

საბიუჯეტო-საგადასახადო სისტემა წამოადგენს ფინანსურ-ეკონომიკური პოლიტიკის შემადგენელ ნაწილს, და ამ პოლიტიკის ერთ-ერთი რთული ფენომენია. ქვეყნის საბიუჯეტო-საგადასახადო სისტემა უნდა წარმოადგენდეს ცალკეული რეგიონების, მასში შემავალი რაიონების, ქალაქების, დაბების, სოფლების,თემების და ა. შ. დამოუკიდებლობის გარანტს.

საქართველოში მოქმედი საგადასახადო კოდექსით , საგადასახადო შემოსავლები ბიუჯეტებს შორის განაწილებულია შემდეგ დონეების მიხედვით: საქართველოს სახელმწიფო და ადგილობრივი თვითმმართველი ერთეულების ბიუჯეტები. ეკონომიკური დოტაციების განაწილება კი წარმოებს, საქართველოს საბიუჯეტო კოდექსით დადგენილი პროცენტული მაჩვენებლების მიხედვით.

საბიუჯეტო კოდექსის თანახმად, სახელმწიფო ბიუჯეტში სრულად ირიცხება მოგების გადასახადი, დამატებითი ღირებულების გადასახადი, აქციზი და იმპორტის გადასახადი, ხოლო რაც შეეხება საშემოსავლო გადასახადს, ავტონომიური რესპუბლიკების მუნიციპალური ბიუჯეტებისათვის განკუთვნილი საშემოსავლო გადასახადის გარდა, იგი სრულად ირიცხება ქვეყნის სახელმწიფო ბიუჯეტში.

ზემოაღნიშნულიდან გამომდინარე, საქართველოში საბიუჯეტო სისტემის სხვადასხვა დონის ბიუჯეტებს შორის შემოსულობების, მათ შორის საგადასახადო შემოსავლების განაწილების საკითხი სერიოზულ მიდგომას საჭიროებს. მიმაჩნია, რომ იგი სრულად ვერ ასახავს ქვეყნის ადმინისტრაციულ-ტერიტორიული დაყოფის სურათს. ტერიტორიული ერთეულების ბიუჯეტების ცნება ვერ ავლენს რეგიონების, ქალაქების, რაიონების, სოფლების, დაბების და ა. შ. ადგილს საერთო სახელმწიფო ბიუჯეტის შემოსავლებში.

ბიუჯეტებს შორის ურთიერთობების ფორმირებაში მნიშვნელოვანი ადგილი უჭირავს ქვეყნის სატრანსფერო პოლიტიკას.

საბიუჯეტო სისტემის სხვადასხვა დონის ბიუჯეტების შემოსავლების გათანაბრება და ადგილობრივი თვითმმართველობისა და მმართველობის ორგანოების ფუნქციების შესრულების ეფექტიანობას სახელმწიფო უზრუნველყოფს. ხელისუფლება, სახელმწიფო ბიუჯეტიდან ახორცლებს სხვადასხვა დონის ბიუჯეტებისთვის ფულადი დახმარების გადაცემას. ადგილობრივი ბიუჯეტები დამოკიდებულნი არიან ხელისუფლების მიერ მათთვის გამოყოფილ ასიგნებზე, სუბვენციებსა და ტრანსფერებზე.

სახელმწიფო ბიუჯეტიდან, ადგილობრივ თვითმართველ ერთეულებს ყოველწლიურად გამოეყოფათ დახმარების სახით გათანაბრებითი ტრანსფერი, რომლის მიზანია ადგილობრივ თვითმართველობებს ექსკლუზიური უფლებამოსილებების განხორციელების მიზნით გააჩნდეთ გათანაბრებული ფინანსური რესურსები. გათანაბრებითი ტრანსფერის გამოყენების მიზნობრიობას ადგილობრივი თვითმართველობის ორგანოები თვითონ ახორციელებენ.

გათანაბრებითი ტრანსფერი გაიანგარიშება ყველა ადგილობრივი თვითმმართველი ერთეულისს მიხედვით, ქვეყნის საბიუჯეტო კოდექსით შემუშავებული წესისა და ფორმულების გამოყენებით.

მსოფლიო მრავალ ქვეყანაში, სახელმწიფო ბიუჯეტიდან გასაცემი ტრანსფერის მოცულობის დადგენა ხდება ფორმულების მეშვეობით, ხოლო საქართველოში, გათანაბრებითი ტრანსფერის გაანგარიშებასთან დაკავშირებული პრაქტიკა ნაკლოვანებებით ხასიათდებოდა, გასაანგარიშებელი ფორმულები ძირითადადთანხების განაწილებაზე იყვნენ ორიენტირებულნი და არა რეგიონის სოციალურ-ეკონომიკური მდგომარეობის გათანაბრებაზე.[52, 49-50] საქართველოში მოქმედი საბიუჯეტო კოდექსის თანახმად, გათანაბრებითი ტრანსფერის გაანგარიშების საწყის ეტაპზე, ქვეყნის ფინანსთა მინისტრის ინდივიდუალური ადმინისტრაციულ-სამართლებრივი აქტის საფუძველზე განისაზღვრება ადგილობრივი თვითმართველი ერთეულების ბიუჯეტების ხარჯებისა და არაფინანსური აქტივების ზრდის მთლიანი მოცულობა, რომლის სიდიდეც არ უნდა იყოს დასაგებმი საბიუჯეტო წლის ნომინალური მთლიანი შიდა პროდუქტის საპროგნოზო მოცულობის 4 პროცენტზე ნაკლები. მომდევნო ეტაპზე გაიანგარიშება ხარჯებისა და არაფინანსური აქტივების ზრდის ჯამის მოცულობა თითოეული ადგილობრივი თვითმართველიერთეულის ბიუჯეტისათვის.

შემდგომ ეტაპზე გაიანგარიშება, ადგილობრივი თვითმართველი ერთეულის ბიუჯეტის შემოსავლები(გრანტის გარდა).

გათანაბრებითი ტრანსფერი არ გამოეყოფა იმ ადგილობრივ თვითმართველ ერთეულს , რომლის ბიუჯეტის შემოსავლებიც აღემატება ბიუჯეტის ხარჯებისა და არაფინანსური აქტივების ზრდის ჯამს.

მომდევნო ეტაპზე კი კოდექსით განსაზღვრული ფორმულის შესაბამისად გამოიანგარიშება თითოეული ადგილობრივი თვითმართველი ერთეულის ბიუჯეტის გათანაბრებითი ტრანსფერის ოდნობა. [35, კარი IV, თავი XI, მუხლი73] თითოეული ადგილობრივი თვითმართველი ერთეულის ბიუჯეტისათვის გადასაცემი გათანაბრებითი ტრანსფერის ოდენობა გამოიანგარიშება საქართველოს საბიუჯეტო კოდექსითდამტკიცებული ფორმულის მიხედვით:

$$T=E-R$$

სადაც:

T -არის ადგილობრივი თვითმართველი ერთეულის ბიუჯეტისათვის გამოსაყოფი გათანაბრებითი ტრანსფერი;

E-ადგილობრივი თვითმართველი ერთეულის ბიუჯეტის ხარჯებისა და არაფინანსური აქტივების ზრდის ჯამი (თვითმართველი ქალაქებისათვის E გამოისახება როგორც $E_{city}$, ხოლო მუნიციპალიტეტებისათვის E-$E_m$ );

R-არის ადგილობრივი თვითმართველი ერთეულის ბიუჯეტის შემოსავლები (გრანდების გარდა),რომელიც ცალკეული ადგილობრივი თვითმართველი ერთეულის ბიუჯეტისათვის გაიანგარიშება მიმდინარე წლის პროგნოზისა და გასული სამი წლის ფაქტობრივი მაჩვენებლების ტენდენციის მიხედვით.

თითოეული ადგილობრივი თვითმართველი ერთეულის ხარჯებისა და არაფინანსური აქტივების ზრდის ჯამი გამოითვლება შემდეგი ფორმულით: [42]

თვითმართველი ქალაქებისათვის :

$E_{city} = S_{city} * G_{city}\%$,

მუნიციპალიტეტებისათვის:

$E_M = S_M * G_M\%$ სადაც,

$E_{city}$ —არის თვითმართველი ქალაქის ბიუჯეტის ხარჯებისა და არაფინანსური აქტივების ზრდის ჯამი;

$E_m$ — მუნიციპალიტეტის ბიუჯეტის ხარჯებისა და არაფინანსური აქტივების ზრდის ჯამი;

$S_{city}$ — თვითმართველი ქალაქის სტატისტიკური და გათანაბრებითი კოეფიციენტების ხვედრითი წილი თვითმართველი ქალაქების კოეფიციენტების საერთო მოცულობაში;

$S_m$ — მუნიციპალიტეტის სტატისტიკური და გათანაბრებითი კოეფიციენტების ხვედრითი წილი, მუნიციპალიტეტების სტატისტიკური და გათანაბრებითი კოეფიციენტების საერთო მოცულობაში;

G ანუ $G_{city}$ —სა და $G_m$ -ის ჯამი წარმოადგენს ადგილობრივი თვითმართველი ერთეულების ბიუჯეტების ხარჯებისა და არაფინანსური აქტივების ზრდის საერთო მოცულობას, რომელიც განისაზღვრება საქართველოს ფინანსთა მინისტრის ინდივიდუალური ადმინისტრაციულ-სამართლებრივი აქტით.

$G_{city}$ — წარმოადგენს თვითმართველი ქალაქების ბიუჯეტების ხარჯებისა და არაფინანსური აქტივების ზრდის მთლიანი მოცულობას, რომელიც განისაზღვრება ადგილობრივი თვითმართველობის ერთეულების ბიუჯეტების ხარჯებისა და არაფინანსური აქტივების ზრდის მთლიანი მოცულობის 72%-ით.

$G_m$ - მუნიციპალიტეტის ბიუჯეტების ხარჯებისა და არაფინანსური აქტივების ზრდის მთლიანი მოცულობა, განისაზღვრება ადგილობრივი თვითმართველობის

ერთეულების ბიუჯეტების ხარჯებისა და არაფინანსური აქტივების ზრდის მთლიანი მოცულობის 28 %-ით. [42]

გათანაბრებითი ტრანსფერის გარდა ბიუჯეტებისათვის შესაძლოა გამოყოფილი იყოს მიზნობრივი და სპეციალური ტრანსფერი. მიზნობრივი ტრანსფერი გამოიყენება დელეგირებული უფლებამოსილებათა განხორციელების მიზნით, ხოლო სპეციალური ტრანსფერი არის საქართველოს საბიუჯეტო სისტემის ერთი დონის ბიუჯეტიდან, მეორე დონის ბიუჯეტისათვის, გათანაბრებითი და მიზნობრივი ტრანსფერის გარდა, ფინანსური დახმარების სახით გამოყოფილი თანხები.

სატრანსფერო პოლიტიკის შემუშავებისას, სახელმწიფო ხელისუფლების მიერ ფინანსური რესურსების განაწილებისას გათვალისწინებული უნდა იყოს სამართლიანობის პრინციპი. რის შედეგადაც დაშვებული იქნება ყველა ერთი დონის თვითმმართველი ერთეულისათვის თანაბარი ოდენობის ტრანსფერების გაცემა.

ყოველივე ზემოაღნიშნულიდან ნათელია, რომ სახელმწიფო განსაზღვრავს ქვეყნის ფულად-საკრედიტო და საბიუჯეტო-საგადასახადო პოლიტიკას, უზრუნველყოფს მისი შესრულებისათვის აუცილებელი კონტროლის მექანიზმის შემუშავებასა და დანერგვას. შესაბამისად გადასახადების დადგენისა და აკრების ყველა ფუნქცია მთლიანად მის ხელშია .

დაბეგვრის მეთოდის მიხედვით მიღებულია გადასახადების დაყოფა ოთხ ჯგუფად:

პროპორციული – ასეთ შემთხვევაში ყველა გადასახადის გადამხდელი იხდის გადასახადს თანაბარი განაკვეთით. საგადასახადო დასაბეგრი ბაზის ზრდის შემთხვევაში გადასახადის განაკვეთი რჩება უცვლელი, ხოლო გადასახადის თანხა იზრდება პროპორციულად. პრინციპი მდგომარეობს იმაში, რომ ყველა ვალდებულია სახელმწიფოს გადაუხადოს თავისი შემოსავლებიდან თანაბარი წილი. მაგალითად საშემოსავლო გადასახადი საქართველოში - 20 %;

პროგრესული – ასეთ შემთხვევაში საგადასახადო ბაზის ზრდის პარალელურად იზრდება საგადასახადო განაკვეთი. აღნიშნული დაბეგვრის პრინციპი მდგომარეობს შეფარდებით პროპორციულ საგადასახადო ტვირში, რათა შემცირდეს უთანასწორობა. ანუ, მაღალი შემოსავლის მქონე გადამხდელი უნდა დაიბეგროს შედარებით უფრო მეტად, ვიდრე დაბალი შემოსავლის მქონე გადამხდელი. მაგალითად, საქართველოში ფიზიკური პირების მიერ ქონების გადასახადის ვალდებულების განსაზღვრისას (გარდა მიწისა));

რეგრესული - აღნიშნული მეთოდი წარმოადგენს პროგრესიული მეთოდის საპირისპიროს, რა დროსაც საგადასახადო ბაზის ზრდასთან ერთად წარმოებს გადასახადის განაკვეთის შემცირება;

ფიქსირებული - აღნიშნულ შემთხვევაში გადასახადის გადამხდელთა საქმიანობის სახეების მიხედვით დგინდება ფიქსირებული თანხა. მაგალითად, საქართველოში თონეში წარმოებული საქონლის მიწოდება, ავტომობილების ტექნიკური მომსახურება ან სოლარიუმის მომსახურება. [24-25]

## 2. საგადასახადო ადმინისტრირების როლი და ადგილი საბიუჯეტო შემოსავლების ფორმირებაში

### 2.1 საგადასახადო ადმინისტრირების პრობლემები საქართველოში თანამედროვე ეტაპზე

გადასახადების ადმინისტრირების პროცესი ნებისმიერი ქვეყნისთვის უმნიშვნელოვანეს საკითხს წარმოადგენს. ის ერთ-ერთია იმ ფაქტორებს შორის, რომელიც საგადასახადო-საბიუჯეტო სისტემის სრულყოფაზე მოქმედებს. სახელმწიფო ბიუჯეტის შევსება და ქვეყნისთვის მნიშვნელოვანი ფინანსური რესურსებით უზრუნველყოფა სწორედ საგადასახადო ადმინისტრირებაზეა დამოკიდებული. ამიტომ გადასახადის გადამხდელის საქმიანობაზე კონტროლის ეფექტიანი სისტემის დანერგვა, ყოველთვის სრულყოფს საგადასახადო კანონმდებლობას.

საქართველოს საგადასახადო კოდექსის თანახმად, გადასახადების ადმინისტრირება ეს არის, გადასახადების გამოანგარიშებასთან, გადახდასთან და დეკლარირებასთან, საგადასახადო კონტროლთან, აგრეთვე გადასახადის გადამხდელთა აღრიცხვასთან, ინფორმირებასთან და საგადასახადო ვალდებულებათა შესრულების უზრუნველყოფასთან დაკავშირებული ფორმების, მეთოდების და წესების ერთობლიობა, რომლებსაც საგადასახადო ორგანოები ახორციელებენ საქართველოს საგადასახადო კანონმდებლობის აღსრულების პროცესში[40, თავი II, მუხლი 8].

გადასახადების ადმინისტრირების და გადახდის სისტემის დარეგულირებისთვის პირველი ნაბიჯები გადაიდგა 1997 წელს მიღებული საგადასახადო კოდექსით, რომლითაც განისაზღვრა კანონის მოქმედების სფერო, გადასახადის განაკვეთები და სახეები, გადახდის პირობები, დეკლარაციის ვადები, გადასახადის გადამხდელის უფლება მოვალეობები . აღსანიშნავია, რომ კოდექსში კონკრეტულად არ იყო გაწერილი თუ რა სახის კონტროლისა და უზრუნველყოფის მექანიზმების არსებობდა გადამხდელის მიმართ. შესაბამისად აუცილებელი გახდა მოქმედი საგადასახადო კოდექსის დახვეწა, რამაც ჯერ არსებულის ცვლილება, ხოლო შემდგომ სრულად გაუქმება და ახალი საგადასახადო კოდექსის მიღება გამოიწვია. [32, გვ 42] ახალი საგადასახადო კოდექსი მიღებული იქნა 2004 წლის 22 დეკემბერს და ამოქმედდა 2005 წლის 1 იანვრიდან, რამაც ფისკალურ პოლიტიკაში მნიშვნელოვანი ცვლილებები შეიტანა.

განხორციელებული ცვლილებების ძირითადი ნაწილი ეხებოდა გადასახადების სახეებისა და მათი განაკვეთების შემცირებას. გადასახადების რაოდენობა 21-დან 7-მდე შემცირდა. ძველისგან განსხვავებით ახალ კოდექსში დეტალურად გაიწერა გადასახადის გადამხდელის უფლება-მოვალეობები, კონტროლისა და უზრუნველყოფის მექანიზმები, გადასახადის ადმინისტრირების წესების და სხვა. (**იხ. დანართი 3**) [32, გვ 42]

| გადასახადები და განაკვეთები 2005 წლის 1 იანვრამდე | გადასახადები და განაკვეთები 2008 წლის 1 იანვრამდე | გადასახადები და განაკვეთები 2008 წლიდან |
|---|---|---|
| საშემოსავლო გადასახადი – 12-20% | საშემოსავლო გადასახადი – 12% | საშემოსავლო გადასახადი – 20% |
| მოგების გადასახადი – 20 % | მოგების გადასახადი – 20% | მოგების გადასახადი – 15% |
| სოციალური დაზღვევის გადასახადი – 31% + 1% | სოციალური გადასახადი – 20% | |
| დღგ – 20% | დღგ – 18% | დღგ – 18% |
| აქციზი – დიფერენცირებული | აქციზი – დიფერენცირებული | აქციზი – დიფერენცირებული |
| ქონების გადასახადი – 0.1 - 1% | ქონების გადასახადი – არა უმეტეს 1% | ქონების გადასახადი – არა უმეტეს 1% |
| ავტოსატრანსპორტო საშუალებათა მესაკუთრეთა გადასახადი - დიფერენცირებული | საბაჟო გადასახადი - დიფერენცირებული | საბაჟო გადასახადი - დიფერენცირებული |
| გადასახადი ქონების გადაცემისთვის | გაუქმებულია | გაუქმებულია |
| გადასახადი ბუნებრივი რესურსებით სარგებლობისთვის | გაუქმებულია | გაუქმებულია |
| გადასახადი მავნე ნივთიერებებით გარემოს დაბინძურებისთვის | გაუქმებულია | გაუქმებულია |
| გადასახადი საქართველოში ავტოსატრანსპორტო საშუალებების შემოსვლისთვის და ზენორმატიული დატვირთვისთვის - დიფერენცირებული | გაუქმებულია | გაუქმებულია |
| ფიქსირებული გადასახადი - დიფერენცირებული | გაუქმებულია | გაუქმებულია |
| მცირე ბიზნესის გადასახადი – 5% | გაუქმებულია | გაუქმებულია |
| ადგილობრივი გადასახადი, რომელიც აერთიანებდა 7 გადასახადს - დიფერენცირებული | გაუქმებულია | გაუქმებულია |

    2008 წლის 1 იანვრიდან ამოქმედდა განახლებული განაკვეთები და ასევე შეიცვალა გადასახადის სახეები, კოდექსიდან ამოღებული იქნა სოციალური გადასახადი, მოგების გადასახადის განაკვეთი 20%-ის ნაცვლად გახდა 15%, ხოლო საშემოსავლო გადასახადი -25 %.

2011 წლიდან ძალაში შევიდა ახალი საგადასახადო კოდექსი, რომელმაც გააერთიანა საბაჟო და საგადასახადო კოდექსები. მიღებული კოდექსის მიზანი მდგომარეობდა, ხელი შეეწყო ბიზნესის სწრაფი ზრდისთვის, გამარტივებული ყოფილიყო საბაჟო და საგადასახადო პროცედურები, რაც თავის მხრივ უზრუნველყოფდა მეწარმეებისთვის საგადასახადო ადმინისტრირების პროცესში აუცილებელი დროის შემცირებას და პროცედურების გამარტივებას.

ზოგადად საგადასახადო ადმინისტრირების მარეგულირებელი აქტები ხშირ შემთხვევაში შედეგების წინასწარი განსაზღვრის გარეშე მიიღება, და ხარვეზების აღმოჩენა უკვე განხორციელებულ ცვლილებებში ხდება. ამიტომ ხშირია კოდექსში ცვლილებები შეტანის შემთხვევები, რაც ართულებს მეწარმეების დამოკიდებულებას გადასახადის გადახდასთან და საგადასახადო ორგანოებთან ურთიერთობასთან მიმართებაში. მაგალითისთვის, 2010 წლის 17 სექტემბერს მიღებული საგადასახადო კოდექსში, რომელიც 2011 წლის 1 იანვარს უნდა ამოქმედებულიყო, ამოქმედებამდე შვიდი საკანონმდებლო ცვლილება განხორციელდა, ხოლო ძალაში შესვლიდან ერთი წლის განმავლობაში 25 ცვლილება იქნა შეტანილი. აქედან გამომდინარე, საგადასახადო სამართლებრივ აქტებში ხშირად შეტანილ ცვლილებებში ფიქსირდება ხარვეზები, რაც დამატებითი ცვლილებების განხორციელების საჭიროებას წარმოშობს. სწორედ საგადასახადო სისტემის ასეთი არამდგრადობა და გაუმართაობა ახდენს გავლენას საგადასახადო სისტემის მდგომარეობაზე საქართველოში.

ამჟამად მოქმედი კოდექსით განსაზღვრულია 6 სახის გადასახადი, რომელთაგან 5 საერთო-სახელმწიფოებრივი ხასიათისა, ხოლო დარჩენილი 1 კი- ადგილობრივ გადასახადს მიეკუთვნება. (იხ. დანართი 4) [32, გვ 43]

| № | გადასახადის დასახელება | გადასახადის განაკვეთი | გადასახადის სახე |
|---|---|---|---|
| 1 | საშემოსავლო გადასახადი | 20% | საერთო - სახელმწიფოებრივი |
| 2 | მოგების გადასახადი | 15% | საერთო - სახელმწიფოებრივი |
| 3 | აქციზი | დიფერენცირებული | საერთო - სახელმწიფოებრივი |
| 4 | დღგ | 18% | საერთო - სახელმწიფოებრივი |
| 5 | საბაჟო გადასახადი | დიფერენცირებული | საერთო - სახელმწიფოებრივი |
| 6 | ქონების გადასახადი | არა უმეტეს 1% | ადგილობრივი |

გადასახადების გაანგარიშების მეთოდები და ადმინისტრირების ფორმები სხვადასხვაა. საერთო-სახელმწიფოებრივი გადასახადის გადახდა აუცილებელია ქვეყნის მთელ ტერიტორიაზე, ხოლო ადგილობრივი გადასახადის გადახდა მხოლოდ ადგილობრივ თვითმმართველ ერთეულში. საქართველოში მოქმედი კანონმდებლობით დადგენილი გადასახადის გადახდა ბიუჯეტში ხორციელდება მხოლოდ ფულადი ფორმით, რომელსაც იხდის გადასახადის გადამხდელი.

საქართველოს ბიუჯეტის შემოსავლების ფორმირებაში მნიშვნელოვანი ადგილი საშემოსავლო და მოგების გადასახადებს უჭირავს. საინტერესოა ბიუჯეტში გადასახდელი თანხის გამოანგარიშებისა და ადმინისტრირების პროცესთან დაკავშირებული დეტალების ანალიზი.

საშემოსავლო გადასახადის გადამხდელად გვევლინებიან რეზიდენტი და არარეზიდენტი ფიზიკური პირები. საშემოსავლო მიზნებისთვის დაბეგვრის ობიექტია:

- რეზიდენტი პირის დასაბეგრი შემოსავალი, რომელიც განისაზღვრება, როგორც სხვაობა კალენდარული წლის ერთობლივ შემოსავალსა და ამ პერიოდისათვის ამ ინსტრუქციით გათვალისწინებული გამოქვითების თანხას შორის.
- არარეზიდენტი ფიზიკური პირის დასაბეგრი შემოსავალი, რომელიც საქმიანობას ეწევა საქართველოში მუდმივი დაწესებულების მეშვეობით, წარმოადგენს საშემოსავლო გადასახადის გადამხდელს დასაბეგრი შემოსავლებიდან, რომელიც განისაზღვრება როგორც სხვაობა კალენდარული წლის განმავლობაში მუდმივ დაწესებულებასთან დაკავშირებული საქართველოში არსებული წყაროებიდან მიღებულ ერთობლივ შემოსავლესა და ამ პერიოდისათვის ამ ინსტრუქციით გათვალისწინებული გამოქვითების თანხებს შორის [40].

აღსანიშნავია, რომ ფიზიკური პირების შემოსავლების დაბეგვრას მოსახლეობის გადასახადებით დაბეგვრის სისტემაში ერთ-ერთი ძირითადი ადგილი ეთმობა. 2004 წლამდე მოქმედი საგადასახადო კოდექსი საშემოსავლო გადასახადის პროგრესულ სისტემას ითვალისწინებდა, დაწესებული იყო 4 განაკვეთი, კერძოდ, 200 ლარამდე შემოსავალზე 12%-ის, 201-დან 359 ლარამდე შემოსავალზე 15%-ის, 351-დან 600 ლარამდე შემოსავალზე17%-ის, ხოლო 601 ლარს ზევით შემოსავლებზე 20%-ის ოდენობით. 2004 წლის საგადასახადო კოდექსით საშემოსავლო გადასახადით დაბეგვრის პროპორციული სისტემა და 12%-იანი განაკვეთი დაკანონდა.

2007 წელს საგადასახადო კოდექსში შევიდა ცვლილებები, რომრლიც ობიექტური მიზეზებით იყო გამოწვეული. ქვეყანაში ეკონომიკური ზრდის მაღალი ტემპის შენარჩუნებისთვის და ინვესტორებისა და ექსპორტის წახალისებისთვის მნიშვნელოვან ეტაპად პირდაპირი გადასახადების განაკვეთების შემცირება ჩაითვალა. საშემოსავლო გადასახადის 25%-იანი განაკვეთი 2009 წლის 1 იანვრიდან 20%-მდე შემცირდა [47, გვ100].

როგორც ცნობილია, შემოსავლების დაბეგვრისას საქართველოს საგადასახადო კანონმდებლობა ითვალისწინებს პროპორციული განაკვეთების გამოყენებას. ერთის მხრივ გამარტივდა საგადასახადო ადმინისტრირების პროცესი, მაგრამ, მეორეს მხრივ ერთნაირ საგადასახადო პირობებში ჩააყენა მაღალშემოსავლიანი, საშუალო და დაბალშემოსავლიანი გადამხდელები.

საგადასახადო ადმინისტრირების კუთხით უნდა იქნეს შემოდებული პროგრესული დაბეგვრის მეთოდი, რაც დადებითად აისახება ფიზიკური პირების დაბეგვრის მექანიზმზე. კონკრეტულად კი საშემოსავლო გადასახადის შემთხვევაში, ე.ი საშემოსავლო გადასახადი უნდა გაიზარდოს შემოსავლების ზრდასთან ერთად. დღესდღეობით საქართველოში ასეთი პრაქტიკა არ არსებობს და საარსებო მინიმუმის ზემოთ მყოფი მოსახლეობა განურჩევლად ყოველთვიური სახელფასო შემოსავლის სიდიდისა, იხდის მის 20%-ს. ეს არ არის სამართლიანი ნორმა, რადგან სხვადასხვა შემოსავლის ადამიანები არ შეიძლება იბეგრებოდნენ თანაბა საპროცენტო განაკვეთით, რადგან ასეთ დროს საგადასახდო ტვირთი და ცხოვრების პრობლემები მცირე შემოსავლიან ფიზიკურ პირს გაცილებით მეტი აქვს.

მსოფლიოს სხვადასხვა ქვეყანაში მოქმედებს პროგრესული დაბეგვრის სისტემა, ანუ რაც მეტია შემოსავალი იმდიან მეტ საშემოსავლოს. მაგალითად აშშ-ში მოქმედებს ორი განაკვეთი, 15% და 39,6%.

საქართველოს სტატისტიკის ეროვნული საგენტოს მონაცემებით, 2007-2011 წლებში დაქირავებით მომუშავეების საშუალოდ ყოველწლიურად 1,60 ათასი კაცით მომატების პირობებში, ყოველთვიური ხელფასი იზრდებოდა 1,2-ჯერ, ხოლო საშემოსავლო გადასახადის მაჩვენებელი 1,3-ჯერ. ამ შედეგების მიხედვით სულ უფრო აქტუალური ხდება დაქირავებით მომუშავე ადამიანების ხელფასის დონის მიხედვით საშემოსავლო გადასახადის საგადასახდო განაკვეთის დიფერენციაციის საკითხი [22 , გვ 105].

საქართველოს საგადასახადო სისტემაში ასევე მნიშვნელოვანი ადგილი უკავია მოგების გადასახადს. საწარმოსთვის დასაბეგრი მოგება განისაზღვრება,როგორც სხვაობა გადასახადის გადამხდელის ერთობლივ შემოსავალსა და საგადასახადო კოდექსით გათვალისწინებული გამოქვითვების თანხებს შორის. მოგების გადასახადს საშემოსავლო გადასახადიდან განასხვავებს ის, რომ გადასახადის გადამხდელია საწარმო, ხოლო საშემოსავლო გადასახადის გადამხდელად გვევლინება ფიზიკური პირი.

საქართველოს საგადასახადო კოდექსის სრულყოფის პირობებში, გამუდმებით მიმდინარეობდა მუშაობა მისი ოპტიმიზაციისათვის, კერძოდ, აღნიშნული გადასახადით დაბეგვრის ობიექტის სწორად განსაზღვრის, საგადასახადო განაკვეთის შემცირებისა და საგადასახადო შეღავათების მოქნილად გამოყენებისთვის, ეფექტურობის მისაღწევად.

2004 წლის საგადასახადო კოდექსით შემოდებული იქნა ძირითადი საშუალებების 100%-იანი გამოქვითვა. გადამხდელებს საშუალება მიეცათ, მოგებას რომელსაც ეკონომიკური საქმიანობიდან იღებდნენ მოგების გადასახადის გადაუხდელად მთლიანად გამოეყენებიათ ძირითადი საშუალებების შესაძენად, წარმოების შემდგომი გაფართოებისათვის მოგების რეინვესტირების გზით. აგრეთვე, ეკონომიკის ზრდისთვის ზრუნვის გამოხატულება იყო მოგების განაკვეთის 20%დან 15%-მდე შემცირება.

2010 წლის საგადასახადო კოდექსით შემოიღეს მიკრო და მცირე მეწარმეობის ცნებები. 30 000 ლარზე ნაკლების წლიური ბრუნვის მქონე მეწარმე, რომელიც ეკონომიკურ საქმიანობაში დაქირავებულ ფიზიკურ პირს არ იყენებს, მიკრო ბიზნესის სტატუსი ენიჭებოდა და მას არანაირი გადასახადის გადახდა არუწევდა. ხოლო მეწარმე ფიზიკური პირი, რომლის წლიური ბრუნვა 100 000 ლარამდეა, ენიჭებოდა მცირე მეწარმის სტატუსი, და ბრუნვიდან გადაიხდიდა 3 %-ს თუ ეკონომიკური საქმიანობიდან მიღებული შემოსავლების 60%-ზე ხარჯების დამადასტურებელი დოკუმენტები გააჩნია, წინააღმდეგ შემთხვევაში ბრუნვიდან 5%-ის გადახდა მოუწევდა.

2017 წლიდან საგადასახადო კოდექსში შევიდა ცვლილებები, რომელიც მოგების გადასახადის ახალი სისტემის შემოღებას ითვალისწინებს. ეს ცვლილებები გულისხმობს მოგების გადასახადის ესტონური მოდელის დანერგვას საქართველოში, რაც ნიშნავს, რომ მოგების გადასახადი მთლიანად გაუქმდა გაუნაწილებელ მოგებაზე და დამფუძნებლის მიერ გატანილი დივიდენდი დაიბეგრება საშემოსავლო გადასახადით, ე.ი. რეინვესტირებული მოგება არ იბეგრება მოგების გადასახადით. ესტონური მოდელი გულისხმობს, რომ ბიზნესის მიერ არ ხდება დადასტურება იმისა, თუ რაში გააკეთა ბიზნესის დამფუძნებელმა რეინვესტირება და ბიზნესი თავისუფალია არჩევანში. გამომდინარე აქედან, მოგების გადასახადით დაბეგვრის ობიექტია:

- რეზიდენტი საწარმოს განაწილებული მოგება;
- გაწეული ხარჯი, რომელიც ეკონომიკურ საქმიანობასთან არ არის დაკავშირებული;

ესტონური მოდელის შემოღების ერთ-ერთი მიზანი მშპ ზრდის ტემპის დაჩქარება, ბიზნესის დაწყებისა გად განვითარებისთვის ხელსაყრელი პირობების შექმნა იყო.

USAID-ის მიერ დაფინანსებული პროექტის ფარგლებში - „მმართველობა განვითარებისთვის" (G4) ჩატარა ესტონური მოდელის მოსალოდნელი შედეგების შეფასება და ანალიზი, კერძოდ კვლევამ აჩვენა, რომ:

- რეფორმას აქვს ინვესტიციების წამახალისებელი ხასიათი და საფონდო კაპიტალი 1,5 წელიწადში 3,23%-ით გაიზრდება;
- დაახლოებით 1,5 წელიწადში მშპ გაიზრდება 1,44 %-ით;
- საერთო კერძო მოხმარება გაიზრდება დაახლოებით 0,85%-ით 1,5 წელიწადში;
- რეფორმა გაზრდის სამთავრობო ბიუჯეტის წლიურ დეფიციტს მაქსიმუმ 3%-ით. ამასთან, საშემოსავლო გადასახადის 1%-ით ზრდა და 1,25%-ით დღგ გაზრდა აღმოფხვრის აღნიშნულ დეფიციტს. აღნიშნული საბიუჯეტო წონასწორობის მისაღწევად მთავრობამ უნდა იფიქროს ხარჯების არგაზრდაზე რეფორმიდან 2-3 წლის განმავლობაში;
- მიმდინარე ანგარიშის დეფიციტი ცოტათი შემცირდება,კომპანიების მიერ დივიდენდების საქართველოსი დატოვებით, გამომდინარე რეფორმის საინვესტიციო ეფექტისა;
- რეფორმის შედეგი დაახლოებით 1,5 წელიწადში გახდება ცნობილი. [5, გვ 155]

რაც შეეხება პრობლემებს, შესაძლებელია ცალკეულ შემთხვევებში თავი იჩინოს გადასახადის გადამხდელთა გარკვეული წრისათვის.

გადასახადის გადამხდელები არარიან მიჩვეულები მოგების გადასახადის ყოველთვიურ დეკლარირებას, რაც მნიშვნელოვან ყურადღებას მოითხოვს ,რომ თვი არიდონ შეცდომებს და შესაბამისად საგადასახადო ჯარიმებს.

რეფორმის შედეგად მოგებით დასაბეგრი ოპერაცია ჩნდება მიკრობიზნესის, ფიქსირებული გადასახადის გადამხდელების და მათი კანონის შესაბამისად შექმნილი საწარმოების საქმიანობაში.

რეფორმის შედეგად გაიზარდა ფინანსურ სტანდარტებზე ცოდნის მოთხოვნა, რაც პრობლემებს შეუქმნის ბიზნესს.

საინტერესოა ბიუჯეტში მოგების გადასახადიდან მისაღები შემოსავლების პროგნოზული შეფასებები 2016-2018 წლებისთვის. შეფასებების თანახმად, წინა წლებთან შედარებით 2018 წლისთვის ამ მაჩვენებლის კლება მოსალოდნელი. აქვე ისიც უნდა აღინიშნოს, რომ ესტონური გამოცდილებიდან გამომდინარე, პირველ ორ წელიწადში დაახლოებით 50%-ის ფარგლებში კლება , ხოლო მომდევნო წლებში მნიშვნელოვანი ზრდა უნდა მოჰყვეს. [6, გვ 124-125] ზემოაღნიშნულის მიუხედავად, რეფორმის ამოქმედების შემდეგაც, მთავრობა არ ამცირებს საბიუჯეტო დანახარჯებს, პირიქით ვალის აღებით და აქციზის გადასახადის ზრდით აპირებს საბიუჯეტო დეფიციტის დაფარვას.  სავარაუდოა, რომ ეკონომიკის ზრდის ტემპი არ გაიზრდება, რადგან მშპ-სთან მიმართებაში მთავრობის ზომა არ შეცვლილა, რაც  ეკონომიკის ზრდას დააჩქარებდა.

რაც შეეხება მოგების გადასახადის განაკვეთს, ის 15%-ს შეადგენს, ხოლო საანგარიშო პერიოდი შეიცვალა და კალენდარული წლის ნაცვლად მოქმედებს კალენდარული თვე, მსგავსად საშემოსავლო გადასახადისა, როცა გადასახადის დეკლარირება და გადახდა ხდება.

საშემოსავლო და მოგების გადასახადის ადმინისტრირების მიზნებისთვის, მნიშვნელოვანია გადასახადის გადამხდელმა სწორად და დროულად აღრიცხოს შემოსავლები და ხარჯები დოკუმენტალურად დადასტურებულ მონაცემთა საფუძველზე და მიაკუთვნოს იმ საანგარიშო პერიოდს, რომელშიც მოხდა მათი მიღება და გაწევა.

დამატებითი ღირებულების გადასახადი, საქართველოს საერთო-სახელმწიფოებრივი გადასახადებიდან ერთ-ერთ ძირითადი გადასახადია. დამატებითი ღირებულების გადასახადიდან შემოსული თანხა საქართველოს ცენტრალურ ბიუჯეტში შედის და ბიუჯეტის საშემოსავლო ნაწილის მნიშვნელოვან წყაროს წარმოადგენს. გადასახადის ძირითადი ფუნქცია ფისკალურია.

საქართველოში დღგ შემოღებულია 1993 წლის 24 დეკემბერს. 1997 წელის საგადასახადო კოდექსით  განისაზღვრა დღგ -ის გადამხდელთა სავალდებულო და ნებაყოფლობითი რეგისტრაციის წესი. თუ პირი ეკონომიკურ საქმიანობას ეწეოდა და უწყვეტი 12 კალენდარული თვის განმავლობაში დღგ-ით დასაბეგრ ოპერაციებს ახორციელებდა და საერთო თანხა 3 000 ლარს აღემატებოდა, ვალდებული ხდებოდა მიემართა საგადასახადო ორგანოსთვის განცხადებით დღგ-ის გადამხდელად რეგისტრაციისათვის. [47, გვ 105] მოგვიანებით, საგადასახადო კოდექსში შეტანილი ცვლილებებით დღგ-ის გადამხდელად გადასახადების გადამხდელთა სავალდებულო რეგისტრაციის ზღვარი 24 000 ლარამდე, ხოლო 2005 წლის 1 იანვრიდან ეს ზღვარი 100 000 ლარს შეადგენს, ხოლო დღგ-ის საგადასახადო განაკვეთი 20-დან 18 %-მდე იქნა შემცირებული.

საქართველოში დამატებითი ღირებულების გადასახადის დაბეგვრის ობიექტს წარმოადგენს: [53, გვ 101]

დასაბეგრი ოპერაცია - საქართველოს ტერიტორიაზე განხორციელებული საქონლის მიწოდება (პირის მიერ სხვა პირისათვის საქონელზე საკუთრების უფლების გადაცემა სასყიდლით ან უსასყიდლოდ) ან/და მომსახურების გაწევა.

იმპორტი - უცხოური საქონლის შემოტანა და საქართველოს საქონლის სტატუსის მინიჭება, რა დროსაც დღგ-ის მიზნებისათვის დასაბეგრი იმპორტის თანხა იანგარიშება, როგორც საბაჟო ღირებულებას დამატებული იმპორტისას გადასახდელი გადასახადები (საქართველოში გადასახდელი დღგ-ის ჩაუთვლელად). თუ საქონელი არის აქციზური ან იბეგრება იმპორტის გადასახადით იმპორტზე, მაშინ გადასახდელი დღგ= (საბაჟო ღირებულებას + იმპორტის გადასახადი+ აქციზი) X 18%.

ექსპორტი, რეექსპორტი - ეროვნული წარმოების სტიმულირების მიზნით აღნიშნულ სასაქონლო ოპერაციებში მოქცეული საქონელი ჩათვლის უფლებით დღგ-გან გათავისუფლებულია. ვინაიდან,არაპირდაპირი გადასახადის საბოლოო გადამხდელი არის მომხმარებელი, ხოლო ექსპორტი/რეექსპორტის შემთხვევაში საქონელი ტოვებს ქვეყნის ტერიტორიას, ეს გადასახადი მძიმე ტვირთად დააწვებოდა ექსპორტიორ მეწარმეს და შეამცირებდა საქონლის კონკურენტუნარიანობას მსოფლიო ბაზარზე. ამიტომ საქართველოდან ექსპორტი არიბეგრება დღგ,ხოლო ამ საქონლის წარმოების პროცესში გადახდილი დღგ კი მწარმოებელი მიიღებს ჩათვლას ან დაბრუნებას.

დროებითი შემოტანა- უცხოური სასქონლის საქართველოს ტერიტორიაზე შემოტანა გარკვეული ვადით სრულად ან ნაწილობრივ გათავისუფლების პირობით. ამ შემთხვევაში საქონლის საბაჟო ტერიტორიაზე ყოფნის ყოველ სრულ და არასრულ კალენდარულ თვეზე გადაიხდება იმ თანხის 0.54%, რომელ თანხასაც იმპორტზე მოქცევისას დაერიცხება დღგ.

აღნიშნული რეგულაცია მნიშვნელოვანია სხვადასხვა ქვეყნის მეწარმეთა საერთაშორისო ეკონომიკური ურთირთობის გაღრმავებისთვის.

2017 წლის 1 იანვრიდან დღგ-ით დაბეგვრას დაექვემდებარა საქონლის მიწოდებამდე ან/და მომსახურების გაწევამდე მყიდველისაგან მიღებული ავანსის თანხები, რა დროსაც დღგ-ით დასაბეგრი ოპერაციის თანხას წარმოადგენს მიღებული ავანსის თანხა დღგ-ს გარეშე, ხოლო დასაბეგრი ოპერაციის განხორციელების დროს ავანსის მიღების მომენტი.

პირის მიერ მისაწოდებელი საქონლის ან/და გასაწევი მომსახურების ღირებულების მხოლოდ ნაწილის, ავანსის სახით, მიღების შემთხვევაში, ავანსის მიღების საანგარიშო პერიოდში დღგ-ით დაბეგვრას დაექვემდებარება მხოლოდ მიღებული ავანსის თანხა, საქონლის /მომსახურების დანარჩენი ნაწილი დღგ-ით დაბეგვრას დაექვემდებარება საქონლის/მომსახურების მიწოდების ან/და მომსახურების გაწევის საანგარიშო პერიოდში. თუ საქონლის და მომსახურების უწყვეტად მიწოდების/გაწევის შემთხვევაა, მაშინ ავანსის თანხების დღგ-ით დაბეგვრის ახალი წესი არ მოქმედებს. [39, მუხლი 161] საგადასახადო ადმინისტრირების მიზნებისთვის, დღგ-ის განაკვეთი შეადგენს დასაბეგრი ბრუნვის ან იმპორტის თანხის 18 %-ს, ხოლო პირი რომელიც რეგისტრირებულია დღგ გადამხდელად ვალდებულია, საგადასახადო ორგანოში წარადგინოს დღგ დეკლარაცია ყოველ სანგარიშო პედიოდზე არაუგვიანეს ამ პერიოდის მომდევნო თვის 15 რიცხვისა და ამავე ვადაში გადაიხადოს დღგ.

დღგ დაბეგვრის შემთხვევაში აუცილებელია გადასახადის დიფერენცირება, და ეკონომიკური პოლიტიკისადმი მისადაგება. აღნიშნული რეგულაცია ხელს შეუწყობს საგადასახადო ბაზის გაფართოებას, ინვესტიციების მაღალი ტემპებით შემოდინებას და აგრეთვე ეკონომიკის პრიორიტეტული დარგების განვითარებას. დღგ დიფერენცირების შედეგად გაიზრდება ბიუჯეტში საგადასახადო შემოსავლები. შესაბამისად დღგ განაკვეთი, რომელიც ამჟამად 18%-ს შეადგენს, საჭიროა შეიცვალოს გარდამავალი ეკონომიკის მდგმარეობის მქონე ქვეყნის ეკონომიკური პოლიტიკის მოთხოვნათა შესაბამისად.

მნიშვნელოვანია დღგ გადასახადით დასაბეგრი ოპერაციების ზღვრული თანხის გაზრდა, რომელიც მოქმედი საგადასახადო კოდექსის მიხედვით 100 000 ლარს შადგენს.

ცვლილების განხორციელების აუცილებლობა გამოწვეულია პირველ რიგში მცირე სამეურნეო სუბიექტების მდგომარეობით, რომელთაგან დღგ-ს ამოღება დიდ სირთულეს წარმოადგენს, რადგან მათ არ აქვთ მოწესრიგებული ბუღალტერია და არ ჰყავთ კვალიფიციური კადრები, რის გამოც ურთულდებათ გადასახდელი თანხის ზუსტი განსაზღვრა. ეს გარემოება კი დაჯარიმების რისკების ზრდას იწვევს. დასმული საკითხის გადაწყვეტა ხელს შეუწყობს მცირე ბიზნესის სტატუსის მქონე მეწარმე ფიზიკურ პირთა რაოდენობის გაზრდას და შესაბამისი ბიზნესექტორის ფორმირების პროცესს. [25, გვ 62] იმპორტის გადასახადი - საბაჟო კოდექსის გაუქმებასთან ერთად, საბაჟო გადასახადს გადაერქვა სახელი და გახდა იმპორტის გადასახადი. იმპორტის გადასახადის ამოცანას ფისკალური ეფექტის მიღებასთან ერთად, იმპორტსა და ექსპორტს, ასევე სავალუტო შემოსავალსა და გასავალს შორის რაციონალური თანაფარდობის მიღწევა წარმოადგენს. მნიშვნელოვანია იმპორტის გადასახადის როლი ეროვნული ეკონომიკის მსოფლიო ეკონომიკაში ინტეგრაციის მიმართულებითაც.

საგადასახადო ურთიერთობათა მოწესრიგების აუცილებლობა კიდევ უფრო აქტუალური გახდა მსოფლიოში საგარეო-სავაჭრო ოპერაციათა ზრდამ, თანამედროვე გლობალური ინფორმაციული ტექმნოლოგიების დანერგვამ, საერთაშორისო ნორმარიულ-სამართლებრივი ბაზის ფორმირებამ და მსოფლიო სავაჭრო ორგანიზაციაში(მსო) გაწევრიანებამ (საქართველო გაწევრიანდა 2000 წლის 14 ივნისს). მსო გაწევრიანების შედეგად:

- საქართველოს საკანონმდებლო ბაზა უფრო ჰარმონიზებული გახდა ევროპილთან, რაც აუცილებელი პირობაა ევროკავშირში ინტეგრირების სტრატეგიული მიზნის მისაღწევად;
- გაღლიერდა საქართველოს ინტეგრაცია მსოფლიო ეკონომიკურ სისტემაში;
- მსო-ს მრავალმხრივი შეთანხმებით საერთაშორისო ბაზარზეგაუმჯობესდა პირობები საქართველოს საექსპორტო პროდუქციისათვის და ქართველი მეწარმეები დაცული გახდნენ ბაზარზე დისკრიმინაციისაგან;

- საქართველომ მიიღო საქონლის და მომსახურებით ვაჭრობის სფეროში წამოჭრილი სადავო საკითხების სამართლიანი და ობიექტური გადაჭრის საშუალება;
- გაუმჯობესდა გარემო უცხოური ინვესტიციებისთვის მოზიდვისთვის.რადგან ინვესტორს ხანგრძლივ პერიოდზე პროგნოზირებადი სავაჭრო პოლიტიკის გარანტიაა გაუჩნდა,ხოლო ქართულ პროდუქციას გაეხსნა საგარეო ბაზრები; [53, გვ 89-90] საქართველოში დღეს არსებული რეგულაციით იმპორტის გადასახადის გადამხდელია პირი , რომელიც საქართველოს საბაჟო საზღვარზე გადაადგილებს საქონელს(გარდა ექსპორტისა),რადროსაც დაბეგვრის ობიექტია ამ საქონლის საბაჟო ღირებულება, რომელიც განისაზღვრება როგორც საქართველოს საბაჟოსაზღვრამდე გაწეული ყველა ხარჯი.

საქართველოში მოქმედებს იმპორტის გადასახადის შემდეგი განაკვეთები:

- 12%-იანი და 5%-იანი განაკვეთი;
- 3%-იანი განაკვეთი, რომელიც მხოლოდ დროებითი შემოტანის ოპერაციაში მოქცეული საქონლის მიმართ გამოიყენება და გამოიანგარიშება იმპორტის გადასახადის იმ მოცულობიდან, რომელიც გადაიხდევინებოდა დროებითი შემოტანის საბაჟო დეკლარაციის რეგისტრაციის დღეს, ამ საქონლის იმპორტში მოქცევისას.

განსხვავებულია სასმელებზე და მსუბუქ ავტომობილებზე დაბეგვრის რეჟიმი, ამ შემთხვევაში იმპორტის გადასახადის ოდენობა არ გამოიანგარიშება დასაბეგრი ობიექტის საბაჟო ღირებულებიდან.

ალკოჰოლურ სასმელებთან დაკავშირებით, 100 ლიტრზე გადასახდელი იმპორტის გადასახადის თანხა შეადგენს იმპორტის გადასახადის განაკვეთის ნამრავლს მოცემულ საქონელში ალკოჰოლის პროცენტული შემცველობის მაჩვენებელზე.

მსუბუქი ავტომობილების იმპორტში მოქცევისას იმპორტის გადასახადის განაკვეთია 0,05 ლარი ავტომობილის ძრავის მოცულობის ყოველ კუბურ სანტიმეტრზე, დამატებული იმპორტის გადასახადის თანხის 5% ავტომობილის ექსპლუატაციაში ყოფნის ყოველი წლისათვის. ეს შეგვიძლია ჩამოვაყალიბოთ შემდეგი ფორმულით:

$$T=(V \times 0.05)+(V \times 0.05) \times 5\% \times Y$$

T-იმპორტის გადასახადი;

V-ძრავის მოცულობა;

Y-ექსპლუატაციაში ყოფნის ვადა;

აუცილებელია სახელმწიფოს მხრიდან ქმედითი ნაბიჯების გადადგმა, რათა იმპორტის გადასახადის მნიშვნელობა წინ იქნეს წამოწეული, რაც მნიშვნელოვანია ეროვნული წარმოების დასაცავად და კონკურენტუნარიანობის ამაღლებისთვის.

იმპორტის გადასახადით დაბეგვრასთან დაკავშირებით გავგაჩნია წინადადებები. საქართველოს ბაზარზე უცხოური საქონლის დაშვების ზედმეტად ლიბერალური პოლიტიკა არსებობს. დღეისათვის საქართველოში მოქმედი იმპორტით დაბეგვრის მექანიზმი ნაკლებად აფერხებს პროდუქციის იმპორტს, ვერ უზრუნველყოფს ეროვნული საწარმოო სექტორის სტიმულირებას , რის გამოც დაუცველი რჩება ეროვნული წარმოება.

ქვეყნის შიდა წარმოების დაცვის კუთხით უფრო დიდი მნიშვნელობა უნდა მიენიჭოს იმპორტის გადასახადს. იმპორტირებული საქონელი სარეოზულ კონკურენციას უწევს ადგილობრივ წარმებას, მაშინ როდესაც საქართველოში დაბალია შრომის მწარმოებლურობა და დაბალია პროდუქციის ფასი. აუილებელია, რომ ქართული პროდუქცია გარკვეულ პერიოდში მაინც პრივილებირებულ რეჟიმში მიმოიქცეოდეს ჩვენს ბაზარზე, ვიდრე მისი იდენტური იმპორტირებული პროდუქცია.

ზოგადად ეკონომიკის დაყრდნობა იმპორტზე ძალიან სარისკოა, რადგან მსოფლიო პოლიტიკა ყოველთვის როდია სტაბილური. იმპორტზე დაფუმვნებული ქვეყნის წინაშე, მათ შორის საქართველოსთვის,დგება საშიშროება, რადგან არაა გამორიცხული იმპორტიორ ქვეყანაში შექმნილმა ეკონომიკურმა კრიზისმა ჩვენს ქვეყანაშიც გამოიწვიოს ფასების ზრდა და საფინანსო სისტემის დესტაბილიზაცია.

შიდა ბაზრის დაცულობისთვის მნიშვნელოვანია სოფლის მეურნეობის მხარდაჭერა, რადგან აღნიშნული სეგმენტი გადასახადის გადამხდელების მხრიდან ყველაზე მაღალ რისკიან სფეროს წარმოადგენს თანხების ინვესტირებისათვის. მოქმედი საგადასახადო კოდექსი აწესებს შეღავათებს აღნიშნული სექტორისთვის, თუმცა იგი უმნიშვნელოა და შესაბამის ეფექტს არ იძლევა.

ქვეყნის საქონლისა და მომსახურების იმპორტი ჯერჯერობით ჭარბობს ექსპორტს. ამასთან მიზანშეწონილი იქნება თუ კვების ძირითად პროდუქტებზე იმპორტის გადასახადების დაწესებისას გათვალისწინებული იქნება ცალკეული პროდუქტების სასურსათო ბალანსების მონაცემები ქვეყნის ადგილობრივი მხარდაჭერისა და აღორძინების მიზნით. გარდა ამისა, იმპორტირებული საქონლის იმპორტის გადასახადის განსაზღვრისას გათვალისწინებული უნდა იყოს ქვეყნის ფისკალური ეფექტი და ეკოლოგიური მდგომარეობა. ამისთვის საჭიროა გამოვიანგარიშოთ ეკოლოგიური კვალის მაჩვენებელი,როგორც საქართველოში წარმოებული პროდუქციის ისე იმპორტირებული პროდუქციის მიხედვით.

აქციზი- აქციზის გადასახადი დღეისათვის გამოიყენება საქონლის და მომსახურების განსაზღვრულ სახეობათა მიმართ და ასრულებს ორ ფუნქციას:
- წარმოადგენს საბიუჯეტო შემოსავლის წყაროს;
- აქციზური საქონლის მოხმარების შეზღუდვისა და საქონლის მოთხოვნა-მიწოდების თანაფარდობის რეგულირების საშუალებაა;

აქციზის გადამხდელია პირი,რომელიც აწარმოებს აქციზურ საქონელს საქართველოში ან ახორციელებს აქციზური საქონლის ექსპორტს/იმპორტს საქართველოში. ახორციელებს ბუნებრივი აიროვანი კონდენსანტის ან/და ბუნებრივი აირის ავტოსატრანსპორტო საშუალებებისთვის მიწოდებას და მობილურ საკომუნიკაციო მომსახურებას.

აქციზით დაბეგვრის ობიექტია: დასაბეგრი ოპერაცია, აქციზური საქონლის ექსპორტი და იმპორტი.

საქართველოში აქციზით დაბეგვრას ექვემდებარება:
- სპირტი და ალკოჰოლური სასმელები;
- თამბაქოს ნაწარმი (თამბაქოს ნედლეულის გარდა);

- მსუბუქი ავტომობილები;
- ბუნებრივი აიროვანი კონდენსატი და ბუნებრივი აირი (გარდა მილსადენისა);
- ნავთობი და ნავთობპროდუქტები, მიღებული ბიტუმოვანი ქანებისაგან(გარდა ნედლი);
- ზეთები და ქვანახშირის ფისებისაგან მაღალ ტემპერატურაზე გამოხდილი პროდუქტები;
- ნავთობის აირები და აირისებრი ნახშირწყალბადები. პიროლიზის თხევადი პროდუქტი;
- მისართი, გამხსნელი, ანტიდეტონატორი;
- საპოხი მასალები და საშუალებები; [40]

რაც შეეხება აქციზური საქონლის საგადასახადო ადმინისტრირების პროცესს, აქციზით დასაბეგრი საქონლის,აქციზური საქონლის იმპორტისა და ექსპორტის თანხა განისაზღვრება:

- ალკოჰოლიანი სასმელისათვის- მოცულობით;
- თამბაქოს ნაწარმისათვის- რაოდენობით ან წონით;
- ნავთობპროდუქტისათვის-წონით(მოცულობით);
- მსუბუქი ავტომობილისათვის- ავტომობილის წლოვანებითა და ძრავის მოცულობით;
- ბუნებრივი აიროვანი კონდენსატისათვის ან/და ბუნებრივი აირისათვის-აირის მოცულობით;
- მობილური საკომუნიკაციო მომსახურების გაწევის შემთხვევაში-მიღებული ან მისაღები კომპენსაციის თანხის მიხედვით.

ადმინისტრირების პროცესში აქციზური საქონლის კონტროლი ერთ-ერთი მნიშვნელოვანი საკითხია. დიდ როლს თამაშობს აქციზური მარკების სისტემა, რომლის ძირითადი ფუნქცია მეწარმეთა კონტროლია. მარკები დამატებითი მაკონტროლებელი მექანიზმია, აქციზური საქონლის უკანონო ბრუნვის წინააღმდეგ. მარკებით ნიშანდებას ექვემდებარება: აქციზით დასაბეგრი ალკოჰოლიანი სასმელები(მათ შორის ლუდი), რომლის ალკოჰოლის შემცველობა აღემატება 1,15 გრადუსს და თამბაქოს ნაწარმი. საგადასახადო ორგანოები დადგენილი წესით ახორციელებენ აქციზური მარკებით სავალდებულო ნიშანდებას დაქვემდებარებული საქონლის ჩამორთმევას.

აქციზის განაკვეთის დაწესების დროს გათვალისწინებული უნდა იყოს ყველა ფაქტორი, რომელიც გარკვეულ ზეგავლენას ახდენს მასზე. გასათვალისწინებელია, რომ განაკვეთებმა და განაკვეთების ზრდამ ხელი არ უნდა შეუწყოს არალეგალური წარმოებისა და იმპორტის ზრდის, კონტრაბანდის და კორუფციის სტიმულირებას. ამავე დროს ყოველმხრივ უნდა იქნეს შესწავლილი მეზობელ ქვეყნებში მოქმედი საგადასახადო განაკვეთები, მოხმარების სპეციფიკა ქვეყნის შიგნით და გადასახადების ადმინისტრირების სტრუქტურა.

საგადასახადო ადმინისტრირების პროცესში, აქციზური საქონლის კონტროლს ერთ-ერთი მნიშვნელოვანი ადგილი უკავია. როგორც ვიცით აქციზური საქონელი იბეგრება ზომის ერთეულის (ლიტრი, კგ, ტონა...) აქციზური განაკვეთით. აღნიშნული პრაქტიკა გამოიყენება ალკოჰოლური სასმელების, სიგარეტის, ნავთობის და სხვა საქონლის წარმოების და შემოტანა-გატანის ოპერაციებისას. მაგრამ საწვავის შემთხვევაში მათი ტონების და ლიტრების მიხედვით აქციზის დაწესება ერთმანეთისგან ვერ მიჯნავს საწვავის სითბური წვის სიმძლავრეს.

აქციზის განაკვეთის გამოანგარიშებისთვის სამართლიანი მეთოდი იქნება, თუ ცალკეულ სასაქონლო ჯგუფებზე მივიღებთ პირობით ზომის ერთეულებს, და შემდეგ მოვახდენთ მასზე აქციზის განაკვეთის დაწესებას.

ქონების გადასახადი- წარმოადგენს საგადასახადო შემოსავლების ერთ-ერთ წყაროს, რა დროსაც ადგილობრივ თვითმმართველობებს უფლება აქვთ საგადასახადო კოდექსით დადგენილი ადგილობრივი გადასახადის თვითმმართველი ერთეულის მთელ ტერიტორიაზე შემოღების შესახებ მიიღონ გადაწყვეტილება.

ქონების გადასახადის მიზნებისთვის დაბეგვრის ობიექტი იყოფა ორ ჯგუფად: დასაბეგრი ქონება და მიწა. ქონების გადასახადის გადამხდელია რეზიდენტი საწარმო/ორგანიზაცია, არარეზიდენტი საწარმო და ფიზიკური პირი.

საწარმოსთვის და ორგანიზაციისთვის ქონების გადასახადის წლიური განაკვეთი განისაზღვრება დასაბეგრი ქონების ღირებულების არა უმეტეს 1 %-ით, რა დროსაც ქონების წლიური ღირებულება გამოიანგარიშება კალენდარული წლის დასაწყისისა და ბოლოსთვის აქტივების საშუალო ღირებულების მიხედვით. შესაბამისად, უნდა აღინიშნოს, რომ ქონების გადასახადის მოცულობა მნიშვნელოვნად გამოხატავს ქვეყანაში ეკონომიკური საქმიანობისათვის გამოყენებული აქტივების ღირებულებას.

ფიზიკური პირის ქონებაზე წლიური გადასახადი დიფერენცირებულია გადამხდელის ოჯახის მიერ საგადასახადო წლის განმავლობაში მიღებული შემოსავლების მიხედვით:

ა) 100 000 ლარამდე შემოსავლისმქონე ოჯახებისთვის-საგადასახადო პერიოდის ბოლოსთვის დასაბეგრი ქონების საბაზრო ღირებულების არანაკლებ 0.05% და არაუმეტეს 0.2 %;

ბ) 100 000 ლარისა და მეტი შემოსავლის მქონე ოჯახებისთვის- საგადასახადო პერიოდის ბოლოსთვის დასაბეგრი ქონების საბაზრო ღირებულების არანაკლებ 0.8 % და არაუმეტეს 1%;

რაც შეეხება სასოფლო-სამეურნეო დანიშნულების მიწაზე გადასახადი გამოიანგარიშება გადასახადის განაკვეთის გამრავლებით მიწის ნაკვეთის ფართობზე(ჰექტრებში).

სასოფლო-სამეურნეო დანიშნულების მიწაზე ქონების გადასახადის წლიური განაკვეთი დიფერენცირებულია ადმინისტრაციული ტერიტორიული ერთეულებისა და მიწის ხარისხის მიხედვით და დგინდება ერთ ჰექტარზე გაანგარიშებით, ლარებში; სასოფლო-სამეურნეო დანიშნულების მიწაზე ქონების გადასახადის განაკვეთები კონკრეტული მიწის ნაკვეთისათვის, მიწის ხარისხისა და მიწის ნაკვეთის ადგილმდებარეობის გათვალისწინებით განისაზღვრება ადგილობრივი თვითმმართველობის წარმომადგენლობითი ორგანოების გადაწყვეტილებით.

არასასოფლო-სამეურნეო დანიშნულების მიწაზე კი ქონების გადასახადის განაკვეთი დგინდება წელიწადში მიწის ერთ კვადრატულ მეტრზე 0.24 ლარის ოდენობით. არასასოფლო-სამეურნეო მიწაზე ქონების გადასახადის განაკვეთები მიწის ნაკვეთის ადგილმდებარეობის გათვალისწინებით, განისაზღვრება ადგილობრივი თვითმმართველობის წარმომადგენლობითი ორგანოების გადაწყვეტილებით, შესაბამისი საბაზრო განაკვეთის გამრავლებით ტერიტორიულ კოეფიციენტზე, ეს კოეფიციენტი არ უნდა იყოს 1,5-ზე მეტი.

2011 წელს საკანონმდებლო ცვლილება შევიდა საგადასახადო კოდექსში, საგადასახადო ადმინისტრირების კუთხით. რის შედეგადაც, ნაცვლად გადასახადის გადამხდელის მთლიანი ქონებისა, ქონების აფასება გავრცელდა მხოლოდ უძრავ ქონებაზე.

საგადასახადო კოდექსში 2017 წლიდან შევიდა ცვლილება,რომლის მიხედვითაც მსუბუქ ავტომობილებზე გადასახადის გადახდა სავალდებულო გახდა და ეს 40 000 ლარზე მეტი ოდენობის წლიური შემოსავლის მქონე ოჯახებს ეხებათ. ქონების გადასახადის წლიური განაკვეთი სხვადასხვა დროს გამოშვებულ ავტომობილებზე განსხვავებულია ავტომობილების ასაკის მიხედვით. 1 წლის ავტომობილზე წლიური გადასახადი შეადგენს 500 ლარს, 1-დან 5 წლამდე ავტომობილზე 240 ლარს, 5-დან 10 წლამდე 120 ლარს, ხოლო 10 წლის ზევით კი 60 ლარს.

ქონების გადასახადის საგადახდო პერიოდია კალენდარული წელი და საწარმო/ორგანიზაცია ქონების გადასახადის დეკლარაციას საგადასახადო ორგანოს წარუდგენს არა უგვიანეს კალენდარული წლის 1 აპრილისა. ქონების გადასახადის გადახდა კი წარმოებს არა უგვიანეს კალენდარული წლის 15 ნოემბრისა.

ქონების გადასახადის ნაწილში საყურადღებოა შემდევი ფაქტი, თუ დასაბეგრი ქონების საბაზრო ფასი აღემატება მის საბალანსო ღირებულებას, მაშინ დასაბეგრი ქონების ფასი განისაზღვრება საბაზრო ფასით და სხვაობაზე დაერიცხება ქონების გადასახადის ძირითადი თანხა. ეს ნორმა საფრთხის მატარებელია, რამდენადაც მიანიშნებს მეწარმეთა ბიზნესინტერესების დაცულობაზე და საგადასახადო ორგანოს უფლებამოსილების საკითხის გადახედვაზე.

სალიზინგო კომპანიისთვის ლიზინგით გაცემულ ქონებაზე განსხვავებული პრაქტიკა არსებობს. ლიზინგის მთელი პერიოდის განმავლობაშ წლიური გადასახადი განისაზღვრება დასაბეგრი ქონების ლიზინგით პირველად გაცემის მომენტისათვის არსებული საბალანსო საწყისი ღირებულების არაუმეტეს 0,6 პროცენტის ოდენობით.

აქედან გამომდინარე ქონებაზე ცვეთის დარიცხვის შედეგად ყოველწლიურად გადასახადის გადამხდელს უმცირდება ქონების გადასახადი, მაგრამ ამავდროულად ლიზინგით გაცემულ ქონებაზე გადასახადის ოდენობა გადაიხდევინება ფიქსირებული ოდენობით. [34]

ქონების გადასახადი ბიზნეს წრეების წარმომადგენლებისთვის ყველაზე არასასურველი გადასახადია. რაც უფრო დიდია საწარმოში ახალი ინვესტიციების დაბანდება, მით უფრო იზრდება მოცემული გადასახადის თანხა. მოქმედი საგადასახადო კოდექსის მიხედვით შედავათებით სარგებლობენ ისეთი დარგები, როგორიცაა: სოფლის მეურნეობა, ტურიზმი, ჯანდაცვა. მაგრამ უმჯობესი იქნება, განსაკუთრებით საქონელმწარმოებელ დარგში, პირველი სამი წლის მანძილზე არ დაიბეგრონ ქონების გადასახადით. ხოლო რაც შეეხება მაღალმთიან რეგიონებს, სასურველია აქ მცხოვრებნი სრულიად განთავისუფლდნენ ქონების გადასახადისაგან.

მიუხედავად იმისა,რომ ბოლო წლების მანძილზე გატარებული ცვლილებებს შედეგად გამარტივდა საგადასახადო კანონმდებლობა და შემცირდა გადასახადების განაკვეთები,საგადასახადო ბაზის სრულყოფა მომავალშიც აუცილებელია. საჭიროა ცვლილებების შესახებ ,როგორც ინვესტორების ისე საზოგადოების ინფორმირება.

ქვეყანაში საგადასახადო სისტემა და პოლიტიკა ისე უნდა მოეწყოს, რომ გადასახადების სახეები და განაკვეთები, საგადასახადო კონტროლი, დავების გადაწყვეტის წესები და საკითხები გადასახადის გადამხდელებსა და სახელმწიფოს შორის ისე უნდა გვარდებოდეს, რომ საფუძველი იყოს მეწარმეობის ხელშეწყობის და კონკურენციის განვითარების, და ზოგადად ვინც სამეწარმეო და ეკონომიკურ საქმიანობას ეწევა თანაბარ პირობებში მოუწიოთ მუშაობა, გამონაკლისის გარეშე.

## 2.2 საგადასახადო ადმინისტრირების გაუმჯობესების მნიშვნელობა საბიუჯეტო შემოსავლების ზრდაში

საგადასახადო ადმინისტრირების ფუნქციონირების მთავარი პირობებია დადგენილი გადასახადების დაბეგვრის ობიექტის, საგნის,ბაზის განსაზღვრა, გამოანგარიშების წესის, გადახდის ვადის და დეკლარირების პერიოდის დადგენა.

საგადასახადო პოლიტიკის ეფექტურობა საფუძველია სახელმწიფოს მიერ ხარჯების გაწევის ვალდებულების შესრულების, ხოლო ბიუჯეტის შევსების ყველაზე მნიშვნელოვან წყაროს გადასახადები წარმოადგენს.

საგადასახადო პოლიტიკამ ერთდროულად უნდა შეასრულოს ფისკალური და მასტიმულირებელი ფუნქცია. რაც თავის მხრივ გულისხმობს, რომ საგადასახადო დაბეგვრის სისტემამ უნდა უზრუნველყოს ეკონომიკის განვითარება, ბაზარზე კონკურენტუნარიანი სუბიექტების არსებობის ხელშეწყობა, ბიუჯეტის საშემოსავლო ნაწილის ფორმირება, დეფიციტის შემცირება და ფინანსური სტაბილიზაციის მიღწევა.

საგადასახადო სისტემის მდგრადობის შესანარჩუნებლად აუცილებელია გადამხდელთა მიერ საგადასახადო კანონმდებლობის დაცვისა და ვალდებულებების შესრულების კონტროლი, რაც საგადასახადო ადმინისტრირების სისტემის მთავარი საქმიანობაა.

საგადასახადო ადმინისტრირებას გააჩნია შემდეგი მახასიათებლები:

- სამართლებრივი უზრუნველყოფა და საკანონმდებლო ბაზის სრულყოფისკენ მისწრაფება;
- სოციალურ პროცესებსა და ბიუჯეტთა შორის ურთიერთობებში მონაწილეობა;
- გადასახადების დროულად და სრულად გადახდაზე კონტროლის განხორციელება, საგადასახადო შემოსავლების პროგნოზირება;
- საგადასახადო ურთიერთობების მართვა, დაბეგვრის სისტემის სრულყოფა ქვეყნის სოციალურ- ეკონომიკური განვითარების მიზნით;
- დაბეგვრის პოლიტიკის ეროვნული და უცხოური გამოცდილების განზოგადება, წესების მკაფიო რეგლამენტირება და დოკუმენტური უზრუნველყოფა.

საგადასახადო ადმინისტრირებისთვის, პროცესის სრულყოფის მიმართულებების შემუშავებისთვის მთავარ პირობას წარმოადგენს ინფორმაციის შეგროვება და ანალიზი. ინფორმაცია მოიცავს საბუღალტრო, საგადასახადო და სტატისტიკური ანგარიშგების სხვადასხვა ფორმებს, ხოლო მისი ანალიზის საფუძველზე კი განისაზღვრება საგადასახადო შემოსავლების ოდენობა გადასახადების სახით, ბიუჯეტებისა და გადამხდელების მიხედვით, აგრეთვე საგადასახადო კონტროლის ეფექტურობა.

ბიუჯეტის წინაშე მდგომი პარამეტრების განსაზღვრისთვის, როგორიცაა რეგიონებისა თუ სხვადასხვა დონის ბიუჯეტების მიხედვით საგადასახადო პოტენციალის განსაზღვრა, საგადასახადო ადმინისტრირება მნიშვნელოვანი მონაწილეა დაგეგმვის და პროგნოზირების კუთხით. დაგეგმვა ხელს უწყობს გადასახადების ფისკალური ფუნქციის შესრულებას და საგადასახადო შეღავათების დაწესების შესაძლებლობას. რაც შეეხება პროგნოზირებას, მისი ძირითადი ამოცანა ფაქტიური საგადასახადო პოტენციალის შეფასებაა, რის საფუძველზეც განისაზღვრება შესაძლო მაქსიმალური ფისკალური ეფექტი და საგადასახადო შემოწმების ობიექტები.

საგადასახადო ვალდებულებების შესრულებისთვის სახელმწიფოს გააჩნია მექანიზმი, რომელიც საგადასახადო რეგულირების სახით არის ცნობილი. იგი გულისხმობს საანგარიშგებო დოკუმენტების შემუშავებას, შეღავათების განსაზღვრას, საგადასახადო ვალდებულების შეუსრულებლობაზე რეაგირებას და სხვა.

რაც შეეხება სახელმწიფოს მხრიდან ფინანსური კონტროლის განხორციელებას, იგი საგადასახადო კონტროლის სახით არის ცნობილი და აერთიანებს სხვადასხვა მეთოდებს, პროცედურებს, მოქმედებებს, საშუალებებს, როგორც საკონტროლო, ისე ანალიტიკური თვალსაზრისით.

სახელმწიფო ბიუჯეტის შემოსულობების ნაწილს გადასახადები შეადგენს. ამიტომ საგადასახადო სისტემის სრულყოფაზე მნიშვნელოვნადაა დამოკიდებული ქვეყნის ეკონომიკური მდგომარეობა. საგადასახადო ადმინისტრირებასთან დაკავშირებული პრობლემების ნაწილი მოგვარდა, ნაწილი კი დღემდე პრობლემატურად რჩება, თუმცა აბსოლიტურად სრულყოფილად ფუნქციონირებადი საგადასახადო სისტემა მსოფლიოს არცერთქვეყანაში არ არსებობს.

ეფექტიანი საგადასახადო პოლიტიკის ფორმირება მრავალ სირთულესთან არის დაკავშირებული, მათ შორი ერთ-ერთი მნიშვნელოვანი გადასახადების ზომის დადგენაა. უნდა შეირჩეს დაბეგვრის ის ზომა, რომელიც ეფექტური იქნება და ხელს შეუწყობს ეკონომიკის განვითარებას და არა პირიქით. ამიტომ საგადასახადო პოლიტიკის ფორმირების მთავრი კრიტერიუმია, რომ ის მარტივი, გასაგები და მისაღები უნდა იყოს საზოგადოების უმეტესი ნაწილისთვის განსაკუთრებით მეწარმეებისთვის. [50]

2008 წლის 1 იანვრიდან საშემოსავლო გადასახადით ფიზიკური პირების შემოსავლები იბეგრებოდა 25 %-ით, მოგვიანებით კი ეს განაკვეთი 20%-მდე შემცირდა, რადგან საშემოსავლო გადასახადმა მოიცვა სოციალური გადასახადი. ამ ცვლილებით მეწარმეები გათავისუფლდნენ გადასახადისაგან, ხოლო გადასახადის სიმძიმე ფიზიკურ პირებს დააწვათ. ჩემი აზრით საშემოსავლო გადასახადთან დაკავშირებული პრობლემების და სადავო საკითხების გადაჭრისთვის მიზანშეწონილია გადასახადის დიფერენციაცია შემოსავლის ოდენობის მიხედვით.

საშემოსავლო გადასახადის მოცულობის ზრდისთვის მნიშვნელოვანია წარმოების სფეროს გამოცოცხლება, რადგან აღნიშნული პოზიტიურად აისახება დასაქმებაზე.

საშემოსავლო და მოგების გადასახადების მობილიზების დონის გაუმჯობესების მიზნით საჭიროა საგადასახადო ადმინისტრირების სისტემის კიდევ უფრო განმტკიცება და დახვეწა, უნდა მოხდეს საგადასახადო კულტურის ამაღლების უზრუნველყოფა. საგადასახადო სამსახურმა კომპლექსური ღონისძიებები უნდა განახორციელოს გადასახადის გადამხდელებზე კონტროლის სისტემატურობის მიღწევის თვალსაზრისით.

საბიუჯეტო შემოსავლების ზრდისთვის და ეფექტიანობის მაქსიმიზაციისთვის პირველ რიგში მნიშვნელოვანია მოგვარდეს ადმინისტრირების გაუმჯობესებასთან დაკავშირებული პრობლემები. ასეთ შემთხვევაში შედეგიც არ დააყოვნებს. ხოლო რაც შეეხება კონკრეტულად დღგ-სგან მიღებულ შემოსავლებს, აუცილებელია მოხდეს გადასახადის დიფერენცირება მსგავსად საშემოსავლო გადასახადისა. რაც შეეხება კანონდარღვევებთან დაკავშირებულ პასუხისმგებლობის საკითხს, გამოყენებული უნდა იყოს ჯარიმებისა და სანქციების მკაცრად განსაზღვრული ნორმები.

კიდევ ერთი წყარო საბიუჯეტო შემოსავლების შევსებისთვის, არის აქციზი. ნავთობპროდუქტების იმპორტის დაბეგვრიდან მიღებული შემოსავლები მნიშვნელოვანი შენატანია ბიუჯეტში. ამ ფაქტის გათვალისწინებით, აუცილებელია გადასახადების ადმინისტრირების გაუმჯობესება რათა, დაწესდეს მკაცრი კონტროლი ნავთობპროდუქტების არალეგალურად შემოტანაზე, დამუშავებასა და შემდგომ რეალიზაციაზე. უნდა მოხდეს ნავთობპროდუქტების მოხმარების ბაზრის საფუძვლიანი გამოკვლევა, რადგან წინასწარ და სწორად იყოს განსაზღვრული საგადასახადო შემოსავლების ოდენობა.

საბაჟო გადასახადი, გადასახადების სახით განსაზღვრული შემოსავლების ნაწილია. ჩვენს ქვეყანაში არსებული საექსპორტო მონაცემები ცხადყოფს, რომ საჭიროა ექსპორტის სასაქონლო სტრუქტურაში არსებული საქონლის მრავალფეროვნების ზრდა, რაც ხელს შეუწყობს საექსპორტო შესაძლებლობების მაქსიმალურად გამოვლენას. რაც შეეხება საბაჟო ტარიფს, მისი დადგენისას სახელმწიფომ მაქსიმალურად უნდა გაითვალისწინოს მეწარმეების ინტერესები და დაიცვას ის. შედეგად მივიღებთ, ადგილობრივი წარმოების განვითარებისთვის ხელშემწყობ პირობებს და თანხებს, რომლებიც საგადასახადო შემოსავლების სახით მიიღება ბიუჯეტში.

არაპირდაპირი გადასახადების ადმინისტრირების გაუმჯობესებას და მოწესრიგებას დიდი ყურადღება უნდა დაეთმოს, რადგან საჭიროა ამ მიმართულებით არსებული პრობლემების გამოსწორება. მართალია არაპირდაპირი გადასახადები სოციალური თვალსაზრისით არასამართლიანია, რადგან შემოსავლების მატება მოსახლეობის ხარჯზე ხდება,მაგრამ სამაგიეროდ მას ფისკალურად უფრო მეტი ეფექტიანობა გააჩნია.

საგადასახადო შემოსავლების ზრდისთვის და ზოგადად ქვეყნის ეკონომიკური აღმავლობისთვის უმნიშვნელოვანესია ქვეყნის პრიორიტეტების გამოკვლევა და დადგენა. საქართველოსთვის პრიორიტეტს წარმოადგენს მრეწველობა და სოფლის მეურნეობა.

სახელმწიფომ თავის მხრივ უნდა იზრუნოს ეროვნული ინტერესების დაცვისათვის, გაატაროს ეფექტური ფულად-საკრედიტო, საბიუჯეტო-საგადასახადო და საერთოდ სწორი ეკონომიკური პოლიტიკა, რამაც თავის მხრივ ხელი უნდა შეუწყოს მრეწველობაში კაპიტალდაბანდებათა გაზრდას. სახელმწიფომ უნდა იზრუნოს საერთაშორისო ეკონომიკური ურთიერთობების გაფართოებაზე და სადაც უპირატესობა მიენიჭება ჩვენი ქვეყნის პროდუქციის საერთაშორისო ბაზარზე რეალიზაციის მხარდაჭერას.

რაც შეეხება სოფლის მეურნეობას და მის განვითარებაზე ზრუნვას. სახელმწიფომ უნდა შეიმუშაოს მკაფიო აგრარული პოლიტიკა და მიიჩნიოს სოფლის მეურნეობა პრიორიტეტად. აგრარული პოლიტიკა ორიენტირებული უნდა იყოს კერძო სექტორზე და ამ კუთხით ინვესტიციების წახალისებაზე. მნიშვნელოვანი ყურადღება უნდა დაეთმოს აგრარული პროდუქციის ექსპორტზე ორიენტაციასაბევრი ობიექტების წრის გაფართოება და ხელსაყრელი ეკონომიკური და პოლიტიკურიც.

მეწარმეობის განვითარებისთვის მნიშვნელოვანია არა მხოლოდ გადასახადების განაკვეთების რეგულირება, არამედ თვით დასაბეგრი ობიექტების წრის გაფართოება და ხელსაყრელი ეკონომიკური და პოლიტიკური გარემოს შექმნა. გამოსწორებას საჭიროებს აღრიცხვა-ანგარიშგებასთან დაკავშირებული პრობლემები.

დაბოლოს უნდა ავღნიშნოთ, რომ ყველა ზემოაღნიშნულთან ერთად საგადასახადო შემოსავლების ზრდა დამოკიდებულია სხვა მრავალ ფაქტორზე,რომელთაგან უმთავრესია:

- დასაბეგრი ობიექტების მოცულობის გაფართოება და ამავდროულად გადამხდელთა რაოდენობის მნიშვნელოვანი ზრდა;
- საგადასახადო სისტემის სრულყოფა;
- საგადასახადო განაკვეთების ოპტიმალური რეგულირება;
- საგადასახადო განაკვეთების სტაბილურობა;

- გადასახადების ამოღების მექანიზმის დახვეწა;
- საგადასახადო სამსახურის მუშაკთა კვალიფიკაციის ამაღლება;
- გადასახადების ამორებასთან დაკავშირებული ხარჯების შემცირება;

ზემოაღნიშნული ღონისძიებები უნდა განხორციელდეს კომპლექსურად, რათა მათი ერთდროულად ამოქმედება გააძლიერებს საგადასახადო სისტემას და თავის მხრივ ხელს შეუწყობს სახელმწიფო ბიუჯეტში საგადასახადო შემოსავლების სახით მობილიზებული თანხების მოცულობის ზრდას. [48-51]

# 3. საბიუჯეტო-საგადასახადო პოლიტიკის უცხოური გამოცდილების დანერგვის პერსპექტივები საქართველოში

## 3.1 საზღვარგარეთის ქვეყნების საგადასახადო ადმინისტრირების სისტემის ანალიზი

საგადასახადო ადმინისტრირება უნდა განვიხილოთ, როგორც ეკონომიკურ საქმიანობაზე სახელმწიფოს ზემოქმედების ძირითადი საშუალება. ნებისმიერი ქვეყნის ძირითად მიზანს უნდა წარმოადგენდეს მმართველობითი გადაწყვეტილებების მიღება და საწარმოთა კონკურენტუნარიანობის ამაღლება.

სხვადასხვა ქვეყნის ადმინისტრირების სისტემები განსხვავებულია, განსხვავებული მოთხოვნების და განვითარების დონის მიხედვით. ისინი იყოფიან სამ ჯგუფად: პირველ ჯგუფს მიეკუთვნება მსხვილი განვითარებული ქვეყნები,რომლებიც რთული საგადასახადო სტრუქტურითა ბიუჯეტში საგადასახადო შემოსავლების დიდი ოდენობით ხასიათდებიან, ასეთებია: აშშ, კანადა, გერმანია, საფრანგეთი, იაპონია, დიდი ბრიტანეთი, იტალია. მეორე ჯგუფს მიეკუთვნებიან ბელგია, ავსტრალია,პორტუგალია, ფინეთი, შვეიცარია, შვედეთი, რომლებიც შედარებით მარტივი საგადასახადო სტრუქტურით ხასიათდებიან. მესამე ჯგუფში შედის მცირე ქვეყნები გადამხდელთა და საგადასახადო მუშაკთა ნაკლები რაოდენობით. მას მიეკუთვნება ლუქსემბურგი, მალტა, კვიპროსი და ა.შ.

მსოფლიოს სხვადასხვა ქვეყნებში საგადასახადო სამსახურების ორგანიზაციული სტრუქტურები განსხვავებულია. მაგალითად ზოგიერთ ქვეყანაში საგადასახადო ურთიერთობების მართვა მკაცრად ცენტრალიზებულია და საგადასახადო სტრუქტურის ყველა რგოლის ზედამხედველობას ახორციელებს ცენტრალური საგადასახადო სამსახური. განსხვავდებიან აგრეთვე საგადასახადო კანონმდებლობის შემუშავება-სრულყოფის მხრივ, საგადასახადო კონტროლის და შემოწმების მექანიზმით.

საგადასახადო კონტროლის მეთოდებში გამოიყოფა, გადამხდელთა რეგისტრაცია და დაბეგვრის ობიექტის აღრიცხვა, ქვეყანაში საგადასახადო კანონმდებლობის დაცვის საერთო კონტროლი და საგადასახადო ანგარიშგების და ვალდებულებების დროულად შესრულების შემოწმება.

ბოლო ეტაპზე, საგადასახადო ორგანოს აუდიტის თანამშრომლები ახორციელებენ კონტროლს, გადამხდელის შესახებ საბანკო საგადასახადო, საბაჟო და სხვა ინფორმაციების მოძიების საფუძველზე.

მსოფლიოს ბევრ ქვეყანაში აუდიტის მიზანი არა სადამსჯელო ღონისძიების გატარება, არამედ სამართალდარღვევის თავიდან აცილებაა.

თანამედროვე განვითარებული ქვეყნების საგადასახადო სამსახურისთვის დამახასიათებელია ფუნქციების გადაცემა კერძო სტრუქტურებისთვის. მაგალითად, კომერციული ბანკების ჩართვა გადასახადების ამოღების მომსახურებაში, დამოუკიდებელ აუდიტორთათვის საგადასახადო კონტროლის გარკვეული სახეების შესრულების მხრივ, საგადასახადო სისტემის კომპიუტერული პროგრამული უზრუნველყოფა კერძო კომპანიების მიერ და სხვა.

განვიხილოთ რამოდენიმე განვითარებული ქვეყნის საგადასახადო ადმინისტრირების თავისებურებები.

ესპანეთში საგადასახადო ადმინისტრირებას ახორციელებს სახელმწიფო საგადასახადო ადმინისტრირების სააგენტო, რომელიც ფინანსთა სამინისტროს დაქვემდებარებაშია. იგი პასუხისმგებელია საგადასახადო და საბაჟო სისტემის მართვაზე, ჩართულია ბიუჯეტის საგადასახადო შემოსავლების მობილიზებაში და სამართალდარღვევების წინააღმდეგ ბრძოლაში.

საგადასახადო ადმინისტრირების სრტუქტურაში შემავალი ინსპექციები ახორციელებენ გადასახადების შეგროვებაზე კონტროლს, რომელიც ორი მეთოდისგან შედგება: ზოგადი კონტროლი, რომელიც მონაცემთა კომპიუტერული დამუშავების გზით ხორციელდება და ადგილობრივი ინსპექციების მონაცემთა ურთიერთშედარება.

გერმანიაში არ არსებობს საგადასახადო სამსახური. არსებობს ფედერალური ფინანსთა სამინისტრო და მიწების ფინანსთა სამინისტროები. ფედერალური ფინანსთა სამინისტრო კონტროლს უწევს აქციზის გადასახადის ადმინისტრირებას, მიწებსა და ფედერაციას შორის ფინანსურ ურთიერთობებს, ხოლო მიწების სამინისტროები ცალკეული გადასახადების ამოღებაზე მუშაობენ.

გერმანიაში ფნანსთა სამინისტროს სპეციალურ ორგანოს წარმოადგენს საგადასახადო პოლიცია, რომელთაც უფლება აქვთ აწარმოონ სამძებრო სამუშაოები, პირადი ჩხრეკა, ექვმიტანილის დაკავება, ნებისმიერი დოკუმენტის ამოღება და სხვა. პრაქტიკულად გააჩნიათ იგივე ფუნქციები რაც კრიმინალურ პოლიციას და ახდნენ ქვეყნის საგადასახადო პოლიტიკის პრაქტიკულ რეალიზაციას. საგადასახადო აუდიტი კი ამოწმებს გადამხდელის მიერ წარმოდგენილ დეკლარაციებს და დარიცხული გადასახადების თანხის სისწორეს.

საგადასახადო კომპეტენციისგან გამოყოფილია და პოლიციის უფლებამოსილებას არის მიკუთვნებული, იმ გადამხდელების გამოვლენა, რომლებიც თავს არიდებენ საგადასახადო ვალდებულებების შესრულებას და სხადიას სხვადასხვა თაღლითურ ქმედებებს.

აშშ გააჩნია მართვის სამი დონე- ფედერალური, შტატების და მუნიციპალური. ყოველ დონეს გააჩნია დაბეგვრის საკუთარი სისტემა და საგადასახადო ორგანოები [45-47].

ქვეყნის შიგნით არსებული გადასახადების ამოღებაზე პასუხისმგებელია ე.წ შიდა შემოსავლების სამსახური, რომელიც ფედერალური გადასახადების გადახდის უზრუნველყოფას ახორციელებს. შიდა შემოსავლების სამსახურს კანონმდებლობის დამრღვევ პირებთან მიმართებაში მინიჭებული აქვს უფლებამოსილება,რომლის თანახმად, მას შეუძლია საგადასახადო დავალიანების ამოიღოს გადამხდელის უძრავ-მოძრავი ქონებით, საბანკო ანგარიშებიდან. ამავდროულად აშშ კანონმდებლობა გადამხდელის ინტერესების დაცვაზეც ზრუნავს. საგადასახადო ორგანოები გამოსცემს და გადამხდელებს სრულიად უსასყიდლოდ მიეწოდება, საგადასახა პროცედურებთან და დოკუმენტაციის მომზადებასთან დაკავშირებული განმარტებითი ბროშრები და ვიდეოკასეტები ვიდეოგანმარტებებით.

საგადასახადო შემოწმების ჩატარების აუცილებლობას განსაზღვრავს კომპიუტერული პროგრამა, რომელიც არჩევს გადამხდელებს გეგმიური შემოწმების ჩასატარებლად. საგადასახადო სამსახურის პოზიცია მდგომარეობს იმაში, რომ თუ საგადასახადო აუდიტი ხშირად განხორციელდება, მაშინ მიღწევა გადამხდელთა მხრიდან საგადასახადო ვალდებულებების სრულად ასახვა დეკლარაციებში. საგადასახადო სამსახური კი გადამხდელებს ავალდებულებს წარმოდგენილ დეკლარაციებში სწორად და სრულად ასახონ დაბეგვრასთან დაკავშირებული ოპერაციები.

საფრანგეთში საგადასახადო სამსახურის სამი დონე არსებობს:

- ცენტრალურ დონეზე ფუნქციონირებს ნაციონალური და საერთაშორისო შემოწმებების სამსახური, ისინი ახორციელებს კონტროლს ისეთ მსხვილ კომპანიებზე, რომლებიც საფრანგეთის საგადასახადო შემოსავლების თითქმის ნახევარს ქმნიან. ესენია ბანკები, საგაზღვევო კომპანიები, კვების მრეწველობის საწარმოები და სხვა.

- ფიზიკური პირების შემოწმებას, რომელთაც აქვთ ძალიან მაღალი შემოსავლები, ახორციელებს საგადასახადო შემოწმების სამსახური. მათ მიეკუთვნება ჟურნალისტები, მსახიობები, სპორტსმენები და სხვა.

- საგადასახადო საგამომძიებო სამსახური ახორციელებს საგადასახადო კონტროლს და ამავდროულად დაკავებულია გადამხდელთა შესახებ ინფორმაციის შეგროვებით.

საფრანგეთში ფუნქციონირებს რეგიონალური და რეგიონთაშორისი საგადასახადო ორგანოები, რომლებიც საშუალო საწარმოთა შემოწმებას აწარმოებენ, ხოლო რაც შეეხება მცირე საწარმოებს, მათ კონტროლს ადგილობრივი საგადასახადო სამსახურები ახორციელებენ.

ყველა გადამხდელი ჩართულია ცენტრალურ მონაცემთა ბაზაში, სადაც იქმნება გადამხდელთა პირადი საქმე, რომელშიც ასახულია სარეგისტრაციო მონაცემები და საბანკო ანგარიშების შესახებ ინფორმაცია. გადამხდელების შესამოწმებლად შერჩევა ხდება მათ მიერ წარმოდგენილი დეკლარაციის საფუძველზე. პირველ რიგში დეკლარაციაში ასახული მონაცემები შედარდება ბაზაში არსებულ ინფორმაციასთან. შეუსაბაობის აღმოჩენის შემთხვევაში ჩატარდება საოფისე კონტროლი,რომლის დროსაც დეტლურადშემოწმდება დეკლარაციაში ასახული ინფორმაცია, გადამხდელის სხვადასხვა ურთიერთობები, კავშირები და შესძლებლობები. საოფისე კონტროლი შეიძლება გახდეს გადამხდელთან საგადასახადო აუდიტის ჩატარების საფუძველი.

საფრანგეთში გადამხდელთა უმეტესი ნაწილი დროულად ასრულებს მათზე დაკისრებულ ვალდებულებებს, შესაბამისად საგადასახადო ორგანო ცდილობს მათთან დაამყაროს კარგი ურთიერთობა. რათქმაუნდა აქაც არსებობს გადამხდელთა გარკვეული კატეგორია, რომლებიც დადგენილ ვადებში არ ასრულებენ მათზე დაკისრებულ მოვალეობებს, აღნიშნული დარღვევები შეიძლება გამოწვეული იყოს გამდამხდელის მიერ დაშვებული უნებლიე შეცდომებით, ამიტომ საგადასახადო ორგანოს თანამშრომლები ცდილობენ ამ ტიპის გადამხდელთა ქცევის შეცვლას და მათთან კარგი ურთიერთობის დამყარებას. ხოლო თუ საგადასახადო აუდიტის დროს გამოვლინდა დარღვევის ნიშნები, ამ შემთხვევაში გამოიყენება რეპრესიული ფუნქცია მაკონტროლებელი ორგანოს მხრიდან.

იაპონიის ფინანსთა სამინისტროს შემადგენლობაში შედის ეროვნული საგადასახადო სამსახური, რომელიც საგადასახადო მართვის და სახელმწიფო კონტროლის ორგანოს წარმოადგენს. ის შედგება რაიონული და ოლქების საგადასახადო სამმართველოებისგან.

იაპონური საგადასახადო სამსახურები ხელმძღვანელობენ შემდეგი პრინციპებით:

- კანონმორჩილი გადამხდელის პატივისცემა;
- საგადასახადო სისტემის სიმარტივე და ხელმისაწვდომობა;
- დასჯასთან შედარებით უპირატესობა გადამხდელთა ნდობას;

საგადასახადო ორგანოთა თანამშრომლები ცდილობენ, რომ გადამხდელებმა გადასახადის გადახდის დროს არ იგრძნონ დისკომფორტი. მაგალითად, საშემოსავლო გადასახადის გადახდელ ფიზიკურ პირებს წინასწარ ეგზავნებათ დეკლარაცია, მასში მითითებული გადამხდელის სახელითა და მისამართით, აგრეთვე იქვე მითითებულია შევსების ინსტრუქცია.

იაპონიის საგადასახადო სისტემა აგებულია არა სასჯელის სიმკაცრეზე, არამედ გადამხდელის კონტროლის სრულყოფილ სისტემაზე. ისინი ცდილობენ დააწესონ გადასახადები,რომლებიც არდააზარალებენ გადამხდელებს და არ აიძულებენ მათ დამალონ შემოსავლები.

საგადასახადო კონტროლის სხვადასხვა მეთოდი არსებობს, რომლის შერჩევაც დამოკიდებულია ბიზნესის ტიპზე და მასშტაბებზე. შესამოწმებელ სიაში ხვდებიან მაღალი შემოსავლის მქონე ფიზიკური პირები და მსხვილი კორპორაციები. შემოწმება შეიძლება ჩატარდეს გადამხდელის საქმიანობის ადგილას ან საგადასახადო ორგანოში [33-52].

## 3.2 საქართველოს და საზღვარგარეთის ქვეყნების საგადასახადო შემოსავლების შედარება და უცხოური გამოცდილების დანერგვის პერსპექტივები

საქართველოს მისწრაფებაა გახდეს ევროკავშირის წევრი ქვეყანა, შესაბამისად სხვა ინსტიტუტების მსგავსად, სახელმწიფო საგადასახადო სისტემაც ევროკავშირის სისტემასთან მისადაგებას მოითხოვს. ასეთი ჰარმონიზების მიზანია მოიხსნას ყოველგვარი ბარიერი შრომითი რესურსების, საქონლის და მომსახურების თავისუფალი მოძრაობისთვის, რომელიც დაუკავშირდება ევროკავშირის სივრცეში თავისუფლად მოძრავ ანალოგიურ რესურსებს. იმისათვის, რომ საქართველო საგადასახადო სისტემის რეფორმების მხრივ დაუახლოვდეს ევროკავშირის წევრი ქვეყნების საგადასახადო სისტემას, საჭიროა ეროვნული საგადასახადო სისტემის სრულყოფა, საგადასახადო ბაზის და სისტემების განგარიშების მეთოდების ევროპულ სივრცესთან დაახლოებადა აქედანვე უნდა დაიწყოს მზადება და იმუშაოს მათი მსგავსი კანონმდებლობით.

საქართველოს საგადასახადო სფეროში განხორციელებული რეფორმები გადასახადების რაოდენობის შემცირების და მთლიანად ტვირთის შემსუბუქების მხრივ, მართალია პასუხობს ევროკავშირის მოთხოვნებს, მაგრამ მათი დაახლოება პირველ რიგში უნდა მოხდეს ევროკავშირის ქვეყნების შესაბამის ანალოგებთან, ასეთი დაახლოება საქართველოს დაეხმარება უცხოური ინვესტიციების მოზიდვაში.

გადასახადების აკრეფა სახელმწიფოს ეკონომიკური ფუნქციის განხორციელების უმნიშვნელოვანესი ბერკეტია და მისი მეშვეობით ფუნქციონირებს სოციალური უზრუნველყოფის სახელმწიფო სისტემა, ხდება სახელმწიფო მმართველობის აპარატის შენახვა.

გადასახადის სახეებსა და მათ განაკვეთებს შორის, როგორც საქართველოში ასევე მთელს მსოფლიოში თვალსაჩინო განსხვავებებია. ეს დამოკიდებულია ქვეყნის ეკონომიკის სპეციფიკაზე, ბიუჯეტის მდგომარეობაზე, განვითარების დონეზე, ამ საკითხისადმი მთავრობის ორგანოების მიდგომაზე და ა.შ. განვიხილოთ საზღვარგარეთის ზოგიერთი ქვეყნის საგადასახადო სისტემების მიმოხილვითი დახასიათებები საქართველოსთან მიმართებაში.

გავაანალიზოთ და შევადაროთ საქართველოში მოქმედი საშემოსავლო გადასახადის განაკვეთი საზღვარგარეთის რამოდენიმე ქვეყნის განაკვეთს. როგორც ვიცით, საქართველოში მოქმედი კანონმდებლობის შესაბამისად მოქმედებს ფიქსირებული 20%-იანი განაკვეთი. შედარებისთვის განვიხილოთ აზერბაიჯანის, ყაზახეთის, გერმანიის, საფრანგეთის და ინგლისის საგადასახადო განაკვეთები. საშემოსავლო გადასახადი აღნიშნულ ქვეყნებში დიფერენცირებულია დასაბეგრი შემოსავლის ოდენობის მიხედვით და 5,5%-45% შორის მერყეობს:

- აზერბაიჯანი - 12%-35%;
  - საფრანგეთი - 5,5%-45%;
  - გერმანია - 14%-45%;
  - ინგლისი - 20-45%;

მონაცემების თანახმად, წარმოდგენილი ქვეყნები მკვეთრად განსხვავდებიან საქართველოსგან განაკვეთების მაჩვენებლით, გადასახადის გაანგარიშების მეთოდებით და ბიუჯეტის მთლიან შემოსავლებში საშემოსავლო გადასახადის ხვედრითი წილით. 2012 წლის მონაცემებით, წარმოდგენილი ქვეყნებიდან ბიუჯეტის შემოსავლებში, საშემოსავლო გადასახადის წილი ყველაზე მაღალი დაფიქსირდა ინგლისში 34.8%, ხოლო ყველაზე დაბალი აზერბაიჯანში 4.4%. საქართველოში საშემოსავლო გადასახადის წილმა 20.4% შეადგინა.

ზოგიერთ ქვეყანაში, განსაზღვრულია იმ მინიმალური შემოსავლის ოდენობაც (დაუბეგრავი მინიმუმი), რომელიც სრულად თავისუფლდება საშემოსავლო გადასახადით დაბეგვრისაგან. კერძოდ, საფრანგეთში საშემოსავლო გადასახადით დასაბეგრი მინიმალური შემოსავალი შეადგენს - 5.963 ევროს, აზერბაიჯანში 1 200 000 მანათს, ხოლო ინგლისში - 9.441 ფუნტ-სტერლინგს. [52, გვ 129-133] საინტერესოა აღნიშნული ქვეყნების მიდგომა მოგების გადასახადის განაკვეთთან მიმართებაში. მოგების გადასახადის 20%-იანი განაკვეთი მოქმედებს აზერბაიჯანში, ყაზახეთში 10%-30% მერყეობს, 15-33.3%-იანი განაკვეთი მოქმედებს საფრანგეთში (მცირე ბიზნესი იბეგრება მოგების 15%-იანი განაკვეთით), ხოლო ინგლისში მოგების გადასახადის განაკვეთი 20-23%-ით განისაზღვრება. გერმანიაში კორპორაციების მოგების გადასახადის მაქსიმალური განაკვეთი – 50%-ია. თუ კორპორაციები მოგებიდან დივიდენდებს გასცემენ, მაშინ – 36%. საშუალოდ კორპორაციის მოგების გადასახადი არ აღემატება – 42%-ს.

საქართველოში მოქმედებს მოგების გადასახადის ყველაზე დაბალი 15%-იანი განაკვეთი (საქართველოში 2017 წლის 1 იანვრიდან მოქმედებს ე.წ ესტონური მოდელი, რაც ნიშნავს, რომ მოგების გადასახადით კომპანია/საწარმო მხოლოდ მოგების განაწილების შემთხვევაში დაიბეგრება, სრული რეინვესტირების შემთხვევაში კი აღნიშნული გადასახადიდან განთავისუფლდება).

დღგ გადასახადით დაბეგვრის განაკვეთი ჩვენს მიერ განხილულ ქვეყნებში 5-30% ფარგლებში მერყეობს. აზერბაიჯანში, საქართველოს მსგავსად მოქმედებს 18%-ანი განაკვეთი მოქმედებს, დასაბეგრი ოპერაციებიდან და დასაბეგრი იმპორტიდან. საფრანგეთში აღნიშნული გადასახადის განაკვეთი 19.1% შეადგენს, მაგრამ აგრეთვე მოქმედებს დიფერენცირებული 7%-იანი, 5.5%-იანი, და 2.1%-იანი განაკვეთები. ყაზახეთში დამატებული ღირებულების გადასახადის დაბეგვრის ობიექტები არის დასაბეგრი ბრუნვა და დასაბეგრი იმპორტი. პირველ შემთხვევაში გადასახადის ამკრეფია საგადასახადო ინსპექცია, მეორე შემთხვევაში – საბაჟო, გადასახადის განაკვეთი 16% შეადგენს. დამატებული ღირებულების გადასახადის განაკვეთი გერმანიაში– 15%-ია (ძირითადი სასურსათო და წიგნადი პროდუქცია იბეგრება 5%-იანი განაკვეთით). ამ გადასახადიდან განთავისუფლებულია საექსპორტო პროდუქცია, სასოფლო-სამეურნეო და ხე-ტყის გადამამუშავებელი მრეწველობის პროდუქცია). დიდ ბრიტანეთში დამატებული ღირებულების გადასახადით ხდება ქვეყნის ბიუჯეტის შემოსავლების 17%-ის მობილიზება. გადასახადის სტანდარტული განაკვეთი ამ ქვეყანაში 17,5%-ია.

საქართველოში დღგ განაკვეთი 2005 წელს 20%-დან 18%-მდე შემცირდა, მიუხედავად ამისა, უნდა აღინიშნოს, რომ წარმოდგენილ ცვლილებას არ გამოუწვევია ბიუჯეტში დღგ-ს სახით მობილიზებული შემოსავლების შემცირება, პირიქით, 2005 წლიდან მოყოლებული სახელმწიფო ბიუჯეტში დღგ-დან მობილიზებული შემოსავლების მაჩვენებელი მკვეთრად და სტაბილურად იზრდება. აქედან გამომდინარე, საქართველოს ხელისუფლებამ კვლავ უნდა განაგრძოს ამ მიმართულებით მუშაობა, რადგან განაკვეთის 18%-დან 15%-მდე შემცირება მოხდეს.

რაც შეეხება აქციზს, აზერბაიჯანში იგი გადაიხდება ყველა აქციზურ საქონელზე, რომელიც იწარმოება ან შემოიზიდება აზერბაიჯანში. აქციზი გადიხდება ნავთობპროდუქტებზე, თამბაქოს ნაწარმზე, საკვებ სპირტზე,ლუდზე და სხვ. გერმანიაში აქციზით იბეგრება მინერალური სათბობი, თამბაქოს ნაწარმი, ყავა, სპირტიანი სასმელები, შამპანური და სხვ. განსხვავებით ზოგიერთი სხვა ქვეყნისგან, გერმანიაში აქციზით იბეგრება სადაზღვევო გარიგებები (12%). ინგლისში აქციზის მოსაკრებლების საშუალო დონე 10-დან 30%-მდე მერყეობს. ყაზახეთში აქციზი გადაიხდება სპირტის ყველა სახეზე, ალკოჰოლურ პროდუქციაზე და თამბაქოზე, მსუბუქ ავტომობილებზე, სათამაშო ბიზნესზე და სხვა.

მიწის გადასახადი გერმანიაში 1,2%-ია. მიწის გადასახადიდან განთავისუფლებული არიან სახელმწიფო საწარმოები და რელიგიური დაწესებულებები. ამასთან, მიწის მეპატრონის შეცვლისას ბიუჯეტში გადაიხდება შესყიდობის ფასის – 2%.

საკუთრებაზე გადასახადი გერმანიაში ამჟამად ფიზიკურ პირებისთვის – 1%-ია, იურიდიული პირებისთვის – 0,6%, ხოლო ქონება რომლის ღირებულება – 120 000 მარკამდე – არ იბეგრება. საფრანგეთში იბეგრება ქონების გაყიდვის ოპერაციები. მაგალითად, უძრავი ქონების მაღალ ფასში გაყიდვა, ფასიანი ქაღალდების მაღალი კურსით გაყიდვა და ა. შ. იბეგრება საკუთრებაც და საკუთრების უფლებაც, ეს გადასახადი პროგრესულია და 0.5%-1.5% შუალედში მერყეობს. ინგლისში ქონების გადასახადს იხდიან მესაკუთრეებიც და არენდატორებიც. გადასახადის განაკვეთს აწესებს ადგილობრივი მუნიციპალიტეტი, ამიტომ მისი სიდიდე საგრაფოების მიხედვით მნიშვნელოვნად განსხვავებულია. ყაზახეთში მიწის გადასახადი დიფერენცირებულია მიწათმოსარგებლის უფლების კატეგორიის მიხედვით: მესაკუთრე და მოსარგებლე. მიწებზე (1 ჰექტარზე) დაწესებულია გადასახადის საბაზო განაკვეთები და დამატებით გადასახადები იმის მიხედვით, თუ რისთვის იყენებენ ამ მიწას, ხოლო ქონების გადასახადს იურიდიული პირები იხდიან 1%-ს, ფიზიკური პირებისთვის კი გამოიყენება პროგრესირებადი გადასახადი. [22, 53-65] ევროპის კავშირში საგადასახადო ბარიერების ლიკვიდაციის საქმეში ძირითად მეთოდად ითვლება წევრ-ქვეყნების საგადასახადო სისტემების თანდათანობით დაახლოება.

ევროკავშირში დაბეგვრის სისტემების ჰარმონიზაცია ემსახურება: თანამეგობრობის ყველა ქვეყნის საგადასახადო სისტემათა სტრუქტურების და გადასახადების ამოღების წესების შესაბამისობაში მოყვანას, ბაზრისთვის გადასახადების ნეიტრალობის უზრუნველყოფას, თანამეგობრობის შიგა საზღვარზე კონტროლის პირობების შექმნას, და გადასახადისგან თავის არიდების შემთხვევების გამორიცხვას.

ჰარმონიზაციის საკითხში განსაკუთრებით წინ წამოიწია დღგ და აქციზის დაბეგვრის უნიფიკაციამ. ჯერჯერობით არ დამდგარა საკითხი ამონაგების ერთიან ორგანებისთვის გადაცემის შესახებ. ევროპული დღგ წარმოადგენს მოხმარების დაბეგვრის უნივერსალურ გადახდას, იგი მოიცავს: ვაჭრობას, მრეწველობას, სოფლის მეურნეობას, მომსახურების გაწევას. წევრ ქვეყნებს კომისიასთან შეთანხმებით ნება ეძლევათ ეკონომიკის ზოგიერთი დარგი დროებით განათავისუფლონ ამ გადასახადისგან და მათი გადასასწყვეტია ასევე, როგორ დაბეგრონ ბრუნვა მცირე ფირმებში, აგრეთვე პირადი მოხმარებისთვის განკუთვნილ მომსახურებაში. დღგ განაკვეთის მინიმალური დონე გამოიყენება ისეთი საქონლის დაბეგვრისას, როგორიცაა: ფარმაცევტული საქონელი, სამედიცინო მოწყობილობები, წიგნები, სამგზავრო ტრანსპორტის მომსახურება და სხვა. არსებობს გამონაკლისები, მაგალითად: ბავშვთა საკვებზე, ტანსაცმელზე და ფეხსაცმელზე დაშვებულია 5%-ზე ნაკლები გადასახადით დაბეგვრა. [4, 119-121] საყურადღებოა ის გარემოება, რომ საქართველოსგან განსხვავებით სოციალური გადასახადი ერთ-ერთი მნიშვნელოვანია განვითარებული ქვეყნების საგადასახადო სისტემაში და საზოგადოების სოციალური დაცვის საქმეში. მნიშვნელოვანია ასევე უცხოეთის ქვეყნებში მიწის დაბეგვრის მექანიზმი. უმეტეს შემთხვევებში მიწა იბეგრება, როგორც უძრავი ქონება, და, საქართველოსაგან განსხვავებით, მას ავტონომიური დაბეგვრის მექანიზმი არ გააჩნია. გერმანიაში, ისევე როგორც საფრანგეთში, მიწა ცალკე არ არის გამოყოფილი ქონებიდან და მისი დაბეგვრა ხორციელდება ქონების ღირებულებიდან, ხოლო გადასახადის განაკვეთი მერყეობს 0.3-0.6% შორის. საქართველოს მსგავსად მიწის გადასახადისაგან განთავისუფლებულია სახელმწიფო ორგანიზაციები, სამედიცინო საქმიანობისთვის, სამეცნიერო-კვლევითი და საღდელ-სასელექციო მიზნისათვის გამოყენებული მიწის ნაკვეთები.

ცალკე აღნიშვნის ღირსია ეკოლოგიური გადასახადის სფეროში ჰარმონიზაციის საკითხი. ამჟამად ევროკავშირის უმრავლეს ქვეყანაში იხდიან ეკოლოგიის გადასახადს. განსაკუთრებით საინტერესოა ენერგო-მატარებლების დაბეგვრა, იბეგრება ნავთობპროდუქტები, ქვანახშირი, ბენზინი. გადასახადის განაკვეთები დამოკიდებულია იმაზე თუ ისინი რამდენად აბინძურებენ გარემოს.

საზღვარგარეთის ქვეყნებში არსებული გამოცდილების და საქართველოში მოქმედი საგადასახადო ადმინისტრირების შედარებითი ანალიზის შედეგად გამოიკვეთა გადასახადების ცალკეული სახეების (საშემოსავლო გადასახადი, მოგების გადასახადი, დამატებული ღირებულების გადასახადი) განაკვეთის დიფერენცირების აუცილებლობა. საგადასახადო სისტემის ლიბერალიზაციის ერთ-ერთი უმტკიცნეულო გზა იქნება მრავალსაფეხურიანი მოგების და საშემოსავლო გადასახადის შემოღება. ამასთანავე, ასეთი მიდგომით შესაძლებელი გახდება პრიორიტეტების სტიმულირება ეკონომიკაში, ადგილობრივი რეალური სექტორის წახალისება, ექსპორტიორი დარგების აღორძინება-განვითარება და ა.შ.

გადასახადების და საგადასახადო ადმინისტრირების ფორმების ცალკეული ქვეყნების ჭრილში გაანალიზებით დავინახეთ, რომ აღნიშნულ სახელმწიფოებში გამოიყენება საშემოსავლო და დღგ გადასახადის დიფერენცირებული განაკვეთები. მათ გააჩნიათ გარკვეული თავისებურებანი და სპეციფიკური ნიშნები, რაც, თავის მხრივ, განპირობებულია ამა თუ იმ ქვეყნის ისტორიული, ეროვნულ – ტრადიციული, ეკონომიკურ – გეოგრაფიული, პოლიტიკური და სხვა თავისებურებებით. მიუხედავად ამისა, საქართველოს საგადასახადო სისტემისა და საზღვარგარეთის ქვეყნების გამოცდილების ანალიზიდან ჩანს, რომ საქართველოსათვის, რომელიც ფაქტობრივად წარმოადგენს განვითარებად ქვეყანას სიღარიბის ზღვარს ქვემოთ მყოფი მოსახლეობის დიდი ხვედრითი წონით, მიზანშეწონილია პროგრესიული განაკვეთების ფართოდ გამოყენება და ამ სფეროში მადლგანვითარებული, ცივილიზებული სახელმწიფოების მდიდარი გამოცდილების გათვალისწინება.

68

## დასკვნა

წინამდებარე ნაშრომში განვიხილეთ საქართველოს საგადასახადო ადმინისტრირების სისტემა და გავაანალიზეთ მისი გავლენა საბიუჯეტო შემოსავლებზე. საკვლევი თემის სრულყოფილი შესწავლისთვის ნაშრომში წარმოვადგინეთ საბიუჯეტო-საგადასახადო სისტემის თავისებურებები, საგადასახადო ადმინისტრირების პრობლემები და მისი გაუმჯობესების გზები, საგადასახადო ადმინისტრირების სისტემის და საბიუჯეტო შემოსავლების ანალიზი საზღვარგარეთის ქვეყნებში, მათი ჰარმონიზაცია საქართველოს საგადასახადო შემოსავლებთან და ა.შ.

სახელმწიფოს ნორმალური ფუნქციონირებისთვის და ეკონომიკის მდგრადი განვითარებისთვის მნიშვნელოვანი როლი ეკუთვნის ეფექტიან საგადასახადო პოლიტიკას. ამიტომ საჭიროა მმართველმა რგოლებმა შექმნან ისეთი პირობები, რაც სამართლებრივ და ეკონომიკურ ბერკეტებში გამოიხატება. საგადასახადო პოლიტიკის გატარებისას არ არსებობს რეცეპტი, სახელმწიფოს საგადასახადო პოლიტიკა უნდა გამომდინარეობდეს კონკრეტული ეკონომიკური სიტუაციიდან და ქვეყნის სტრატეგიული გეგმიდან. გადასახადები არ არის საზოგადოების წინაშე არსებული ყველა პრობლემების გადაჭრის გზა, თუმცა საგადასახადო სისტემას შეუძლია ანტიკრიზისული ღონისძიებების გატარება. იმისათვის, რომ სახელმწიფომ შეძლოს მასზე დაკისრებული ფუნქციების შესრულება, აუცილებელია ქვეყანაში შემოსავლების განაწილების ოპტიმალური გზების პოვნა, რაც ეკონომიკის საქმიანობის სტიმულირების და მოსახლეობის სოციალური დაცვისთვის მნიშვნელოვანია. ამ ყველაფრის ეფექტურად გადაჭრაში უმთავრესია საგადასახადო ადმინისტრირების სისტემის როლი.

საქართველოს საგადასახადო ადმინისტრირების სისტემის ანალიზით შეგვიძლია დავასკვნათ, რომ ის არის რეალურ ეკონომიკის მდგომარეობაზე მორგებული სისტემა, რომელიც ხასიათდება: რთული და არასრულყოფილი საკანონმდებლო ბაზით, მაღალი საგადასახადო სიმძიმით და დაბეგვრის არასტაბილური გარემოთი.

იმისათვის, რომ საგადასახადო ადმინისტრირების სისტემამ ფისკალურ ფუნქციასთან ერთად კანონიერების და გადამხდელთა გადახდისუნარიანობის უზრუნველყოფა მოახდინოს, აუცილებელია სისტემამ უზრუნველყოს გადამხდელთა ინფორმირება და საგადასახადო ვალდებულებების სწორად შესრულებაში დახმარება. კანონმდებლობა უნდა იყოს, როგორც გადამხდელისთვის ისე კონტროლის განმახორციელებლისთვის გასაგებად გამართული, რასაც საქართველოს საგადასახადო კოდექსის შესაბამისად გამოცემულ კანონქვემდებარე აქტებზე ვერ ვიტყვით.

განვითარებული ქვეყნების მიდგომა საგადასახადო ადმინისტრირების განხორციელების მიმართ განსხვავებულია, ისინი ცდილობენ გადამხდელებს შეუქმნან შესაბამისი პირობები, რათა შეასრულონ დაკისრებული ვალდებულებები კანონმდებლობის დაცვით. ამ მიზნით ხდება მათი ინფორმირება, კონსულტაციების გაწევა, ბროშურების დარიგება, ვალდებულებათა შესრულების დროის შეხსენება და სხვა. ისინი უფრთხილდებიან კეთილსინდისიერ გადამხდელებს, რათა მათ თავი არ იგრძნონ არაკომფორტულად, რადგან მათი არსებობა საგადასახადო სისტემის გამართულად მუშაობის მთავარი ბერკეტია.

განვითარებული ქვეყნების საგადასახადო სისტემა საკმაოდ დიდი ხნის განმავლობაში განიცდიდა ცვლილებებს სრულყოფისა და ეფექტიანობის მიღწევისთვის. საბოლოოდ ჩმოყალიბდა საგადასახადო სისტემის დონე, სადაც მაღალია საგადასახადო კულტურა, ეკონომიკის განვითარების სტიმულირება და ა.შ საქართველოში არსებული საგადასახადო შეღავათები, რომლებიც უნდა უზრუნველყოფდეს საშუალო და დაბალშემოსავლიანი გადამხდელების რაოდენობის ზრდას, პირიქით, მორგებულია მსხვილ ბიზნესზე. მდგომარეობის გამოსასწორებლად აუცილებელია პრიორიტეტებად არჩეული უნდა იყოს მცირე და საშუალო მეწარმეები, რადგან მათი განვითარების გარეშე ქვეყანაში ვერ იარსებებს მოსახლეობის საშუალო ფენა.

არაერთხელ ავღნიშნეთ, რომ საქართველოში არის საგადასახადო კულტურის დაბალი დონე, რაც კანონმდებლობის ცოდნის დონის ამაღლებით, კანონმდებლობის დახვეწით, საგადასახადო ტვირთის სამართლიანად გადანაწილებით და კონტროლის სამართლიანი და ეფექტური ღონისძიებების გატარებით უნდა აღმოიფხვრას.

საქართველოში და არა მხოლოდ, შეღავათები მოქმედებს ექსპორტზე და წარმატებით გამოიყენება ბევრ ქვეყანაში, მაგრამ მხოლოდ ექსპორტის წახალისება არ არის საკმარისი, სახელმწიფომ უნდა იზრუნოს შიდა წარმოების შექმნაზე.

სანქციების დადგენის კუთხით, მთავარია ოპტიმალური პირობის დაცვა. საქართველოს საგადასახადო კოდექსით განსაზღვრულმა სანქციებმა ბევრჯერ განიცადა ცვლილება და ჯერ კიდევ არ არის ოპტიმალური ნიშნული ჩამოყალიბებული. ოპტიმალურიბა გულისხმობს, რომ სანქცია არუნდა იყოს არც ისეთი დაბალი,რომ გადამხდელს უდირდეს ვალდებულების შეუსრულებლობა და არც ისეთი მაღალი, რომ ხელი შეუშალოს სუბიექტის შემდგომ საქმიანობას.

ქვეყნის განვითარების ერთ-ერთი მაჩვენებელი სახადასახადო შემოსავლებში, პირდაპირი გადასახადების ხვედრითი წილის ზრდაა, რაც განვითარებული ქვეყნებისთვისაა დამახასიათებელი. საქარველოში კი პირიქითაა, არაპირდაპირი გადასახადების ხვედრითი წილი ჭარბობს. ამიტომ სახელმწიფომ უნდა იზრუნოს პირდაპირი გადასახადების ზრდის ხელშეწყობაზე.

საგადასახადო ადმინისტრირების სისტემის გაჯანსარებისთვის აუცილებელია კომპლექსური მიდგომა, მხოლოდ გადასახადების რაოდენობის შემცირება არ არის საკმარისი, მნიშვნელოვანია რეფორმების გატარების დროს უცხოური გამოცდილების გაზიარება და ეროვნული თავისებურებების გათვალისწინებით მსოფლიოში წარმატებული პროექტების ადაპტირება.

ყოველივე ზემოთ მოყვანილი წინადადებების რეალიზაციის შემთხვევაში მოსალოდნელია, რომ ქვეყანაში გაუმჯობესდეს ბიზნეს გარემო და საგადასახადო ადმინისტრირების პროცესი, საგადასახადო შემოსავლების ხარჯზე გაიზრდება საბიუჯეტო შემოსავლები და ამაღლდება საზოგადოების საგადასახადო კულტურის დონე.